\documentclass[fleqn,usenatbib]{mnras}

\usepackage{newtxtext,newtxmath}

\usepackage[T1]{fontenc}

\DeclareRobustCommand{\VAN}[3]{#2}
\let\VANthebibliography\thebibliography
\def\thebibliography{\DeclareRobustCommand{\VAN}[3]{##3}\VANthebibliography}

\usepackage{graphicx}	
\usepackage{amsmath}	
\usepackage{comment}
\usepackage{enumitem}

\title[Abundant SiO in TWA\,5\,B]{The CRIMSON survey I: super-stellar SiO in the directly imaged companion TWA\,5\,B from high-resolution \textit{M}-band spectroscopy}

\author[L. T. Parker et al.]{Luke T. Parker,$^{1}$\thanks{E-mail: luke.parker@physics.ox.ac.uk}
Jayne L. Birkby,$^{1}$
Siddharth Gandhi,$^{2,3}$
Vivien Parmentier,$^{4}$
Vatsal Panwar,$^{5,2,3}$
 \newauthor
Matteo Brogi,$^{6,7}$ 
and Sophia R. Vaughan $^{8}$  
\\
\\
$^{1}$Astrophysics, Department of Physics, University of Oxford, Denys Wilkinson Building, Keble Road, Oxford, OX1 3RH, UK\\
$^{2}$Department of Physics, University of Warwick, Coventry CV47AL, UK\\
$^{3}$Centre for Exoplanets and Habitability, University of Warwick, Coventry CV47AL, UK\\
$^{4}$Université Côte d’Azur, Observatoire de la Côte d’Azur, CNRS, Laboratoire Lagrange, France\\
$^{5}$University of Birmingham, School of Physics \& Astronomy, Birmingham, B15 2TT, UK\\
$^{6}$Dipartimento di Fisica, Universit\`a degli Studi di Torino, via Pietro Giuria 1, I-10125, Torino, Italy\\
$^{7}$INAF-Osservatorio Astrofisico di Torino, Via Osservatorio 20, I-10025 Pino Torinese, Italy\\
$^{8}$Max-Planck-Institut f\"ur Astronomie, K\"onigstuhl 17, 69117 Heidelberg, Germany\\
}

\date{Accepted 2026 June 1. Received 2026 May 22; in original form 2026 April 10}

\pubyear{2026}

\begin{document}
\label{firstpage}
\pagerange{\pageref{firstpage}--\pageref{lastpage}}
\maketitle

\begin{abstract}

Silicon is a key refractory element in giant planet atmospheres, which governs the formation of magnesium-silicate clouds, and reflects the quantity of silicates accreted during formation. While observations of directly imaged giant exoplanets have focused on the measurement of volatile species (e.g. CO, H$_2$O), high-resolution spectroscopy with CRIRES+ \textit{M}-band provides access to gas phase silicon chemistry in sub-stellar atmospheres, through the ro-vibrational band head of SiO at 4 $\mu$m. Here, we present the first results of the CRIMSON survey of silicon chemistry in directly imaged companions with CRIRES+ \textit{M}-band. We report the strong detection of gaseous SiO (S/N = 7.5) in the directly imaged companion TWA\,5\,B, with an atmospheric abundance of log(SiO) = $-3.56^{+0.42}_{-0.32}$ VMR, providing access to the refractory content of the atmosphere. The high retrieved SiO abundance implies the absence of significant magnesium-silicate cloud condensation, and thus the atmospheric silicon abundance is contained almost entirely within the observed gas phase SiO. Using the detection of refractory silicon, together with strong detections of the volatile species CO (S/N = 9.1) and H$_2$O (S/N = 18.8), we measure a stellar C/O and a marginally sub-stellar O/Si and C/Si, but a super-stellar Si/H ([Si/H]$_{\star}$ = $1.41^{+0.42}_{-0.32}$). Collectively, these volatile-to-refractory ratios are consistent with formation through core-accretion beyond the CO snowline, or gravitational instability followed by substantial solid enrichment. Finally, we discuss how gas phase SiO provides a unique diagnostic of the cloud properties in hot gas-giants, and can be used to probe the dominant cloud species forming across the directly imaged planet and isolated brown dwarf populations.
\end{abstract}




\section{Introduction}
\label{sec:Intro}

Exoplanet discovery surveys have revealed planetary systems with diverse architectures, far beyond our expectations from our own solar system, which challenge our understanding of planet formation \citep{Winn2015}. To understand the mechanisms that drive such a diversity in the exoplanet population, we aim to link the present-day atmospheric composition of these exoplanets to their formation history \citep{Molliere2022, Oberg2011, Feinstein2025}. Previous works have targeted the relative abundance of oxygen and carbon (C/O ratios) from volatile species such as CO and H$_2$O to trace planet formation, under the assumption that the present-day atmospheric C/O ratio reflects the chemical partitioning in the region of the protoplanetary disc in which the planet formed \citep{Oberg2011, Madhusudhan2012}. In this framework, volatile species condense at distinct radii in the disc according to their condensation temperatures (ice lines), and thus the protoplanetary disc chemistry is stratified between gas and solid phases, producing radial variations in the gas-phase C/O. Accordingly, planets that accrete their atmospheres at different disc radii, or migrate across these ice lines, may inherit distinct C/O signatures that encode their formation history \citep{Oberg2011}. However, applying the C/O ratio alone with the classic assumption of a static disc \citep[e.g.][]{Oberg2011} has been shown to lead to degeneracies in the predicted formation positions of planets due to the complex and time-dependent thermal and chemical processing of material in the protoplanetary disc \citep{Mordasini2016, Lichtenberg2021, Oberg2023, Mordasini2024}.

More recently, $^{12}$C/$^{13}$C isotopologue ratios \citep{Molliere2019b, Zhang2021a, Zhang2021b} have been proposed as a means of tracing the ratio of rocks to ices accreted during formation, and have been traced across both directly imaged planets \citep[e.g.][]{deRegt2024, GonzalezPicos2025, Zhang2024, Xuan2024} and short-period hot Jupiters \citep{Smith2024b, Finnerty2024}. However, the robust detection of isotopologues requires high S/N spectra, and further work is required to understand the processes governing the fractionation of isotopes in the primordial disc \citep{Smith2015,Crossfield2019}. A relatively novel complementary tracer of formation histories of giant planets is through observations of refractory material (i.e. material with a high vaporisation temperature such as Si, Fe, and other metals). The observation of refractory species can break the degeneracies implicit within C/O ratios, as these elements remain in the solid phase throughout most of the protoplanetary disc, and thus the present-day atmospheric abundance ratios of refractory elements in exoplanetary atmospheres reflect their ratios during formation \citep{Turrini2021, Schneider2021a, Schneider2021b, Lothringer2021,Chachan2023}. The Si/H abundance ratio traces the ratio of solids-to-gaseous components accreted, while the O/Si and C/Si ratios track the ratio of ices-to-rocks accreted during formation. These constraints offer a pathway to separate the formation location and initial composition of the protoplanetary disc in which a planet formed \citep{Oberg2011, Lothringer2021, Chachan2023}.

In transiting planets, refractory elements have been routinely detected in ultra-hot Jupiters using high-resolution spectroscopy (HRS) at spectral resolutions of R $\approx$ 100$\,$000 (e.g. \citealt{Hoeijmakers2018_nature, Prinoth2022, Gandhi2023_UHJ, Simonnin2025}). These observations rely on the presence of neutral (e.g. Fe, Si, Ti) and ionised (e.g. Ca+, Fe+, Ti+) metals which produce abundant spectral lines, primarily present at visible wavelengths but which additionally pervade into the NIR (e.g. \citealt{vanSluijs2025}). The inclusion of NIR observations of volatile species has permitted the measurement of volatile-to-refractory ratios for the ultra-hot Jupiters WASP-121 b and WASP-189 b \citep{Pelletier2025, Smith2024, Sanchez2025}. However, the detection of these atomic and ionised refractory species is at present limited to the ultra-hot Jupiter population, requiring T$_{\text{eq}} \gtrsim$ 2000 K \citep{Snellen2025}. Space-based spectroscopy provides an alternative route to accessing spectroscopic features from refractory species and observations with HST have probed refractory species in the UV, but can suffer from blended spectral features and degeneracies with stellar activity due to the lower spectral resolution \citep{Lothringer2022, Lothringer2025, Baldwin2026, Chachan2025}. Recently, the detection of VO and SiO in the ultra-hot Jupiter WASP-121 b \citep{Gapp2025, Evans-Soma2025, Pelletier2026} have demonstrated that refractory elements can also be accessed in molecular form in ultra-hot planets with JWST spectroscopy. Additionally, the moderately refractory element sulphur has been detected in molecular form in giant planets with JWST, as SO$_2$ \citep{Alderson2023, Beatty2024, Gressier2025, Crossfield2025} and H$_2$S \citep{Fu2024, Ruffio2026}.

On tidally locked planets the interpretation of the observed abundances of refractory species is complicated by the condensation of refractory elements. The impact of extreme hemispherical heating and inverted atmospheres on hot Jupiters lead to predictions of complex cloud formation of refractory elements (e.g. \citealt{Parmentier2016, Lee2017}). In hot Jupiters the main thermal gradient responsible for cloud formation is the permanent day-to-night temperature contrast. Clouds can therefore form on the cooler nightsides of hot Jupiters, leading to the rain out of refractory material in the deeper, unobservable layers of the atmosphere \citep{Parmentier2013}. This nightside cold trap can deplete the whole atmosphere of specific refractory species (e.g. V, Ti, Si). This includes the depletion of refractory elements on the hotter dayside where no clouds are forming (e.g. \citealt{Pelletier2023}), distorting the observed refractory abundances from transmission and emission spectroscopy. 

In comparison, the cloud properties of the directly imaged planet population are better understood, as they are informed by the detailed study of the analogous isolated brown dwarf population \citep{Faherty2016, Suarez2022, Vos2023, Calamari2026}. The condensation of cloud species containing refractory elements (e.g. magnesium-silicate clouds; MgSiO$_3$, Mg$_2$SiO$_4$) is one of the dominant processes that shapes the population of self-luminous worlds, from the onset of condensation at the M/L transition ($\sim$2200 K) through to the L/T transition ($\sim$1300 K) at which the silicate clouds sink below the photosphere \citep{Ackerman2001, Visscher2010, Teinturier2026}. This physical process shapes the observable properties of these objects through reddening to later L type spectral classes (see Figure \ref{fig:survey_plot}), followed by the rapid breakup into patchy clouds at the L/T transition, driving a shift to blue J-K colours and the onset of significant planetary scale variability \citep{Saumon2008, Apai2013, Tan2019}. Directly imaged planets and low-mass brown dwarf companions follow the same sequence as they cool, but show on average redder colours, with the onset of cloud breakup at the L/T transition occurring at cooler temperatures, suggesting thicker clouds driven by their lower surface gravity \citep{Stephens2009, Faherty2016, Suarez2023}. Refractory species play a vital role in both the formation and the evolution of directly imaged worlds but, due to prohibitive planet-to-star contrast ratios at visible wavelengths, accessing refractory species in this wavelength regime is challenging \citep{Currie2023}. To trace volatile-to-refractory ratios and cloud formation in the directly imaged planet population, we require access to refractory elements in molecular form. Molecular species containing refractory elements have strong near infrared opacities, and are present in planetary atmospheres at cooler temperatures than atomic and ionised species. 

High-resolution spectroscopy in the thermal infrared \textit{M}-band (3.5 - 5.2 $\mu$m\footnote{The \textit{M}-band strictly refers to the atmospheric window between approximately 4.5–5.2 $\mu$m, with the L-band denoting the 2.9-4.1 $\mu$m window. In this work we use an \textit{M}-band grating setting with CRIRES+ spanning 3.5 - 5.2 $\mu$m, which additionally covers red regions of the \textit{L}-band. For simplicity we refer to the 3.5 - 5.2 $\mu$m wavelength region covered in this work as the \textit{M}-band.}) offers a new parameter space for accessing refractory species in molecular form. The \textit{M}-band is a challenging wavelength to observe from the ground, with high thermal background noise and telluric contamination, but the first tests of HRS with CRIRES+ in the \textit{M}-band observed marginal evidence for gaseous SiO at 4 $\mu$m in the archetypal directly imaged planet $\beta$ Pic b \citep{Parker2024, Janson2025}. Gaseous SiO is the near exclusive carrier of atmospheric silicon in hot atmospheres without the condensation of (magnesium-)silicate clouds\footnote{Under equilibrium conditions at $\sim$2000 K the next most abundant silicon-bearing molecule, SiS, holds $\sim$1 per cent of the atmospheric silicon abundance \citep{Visscher2010}.}, and thus the silicon abundance locked in SiO directly traces the initial refractory content of accreted primordial material \citep{Lothringer2021, Lothringer2022}. The observed atmospheric abundance of SiO is additionally a sensitive diagnostic of cloud formation, as the gas phase SiO abundance is readily depleted by the onset of (magnesium-)silicate condensation \citep{Visscher2010, Kitzmann2024}. Here, we present CRIRES+ \textit{M}-band observations of the directly imaged companion TWA\,5\,B. With a temperature of T$_{\text{eq}} \approx$ 2400 K, TWA\,5\,B is ideally situated to detect gas phase SiO, and is predicted to be above the condensation temperature of many cloud condensates that would deplete the atmospheric abundance of refractory material, including silicon. 

Below, we introduce the CRIMSON survey of directly imaged companions, and briefly outline the existing research on the TWA\,5 system. In Section \ref{sec:Observations} we describe our observations and we outline our methods in Section \ref{sec:Methods}. In Sections \ref{sec:Results} and \ref{sec:Discussion}, we present our results and discuss their implications for future studies. We conclude in Section \ref{sec:Conclusions}.

\subsection{The CRIMSON survey}

\begin{figure}
    \includegraphics[width=\columnwidth]{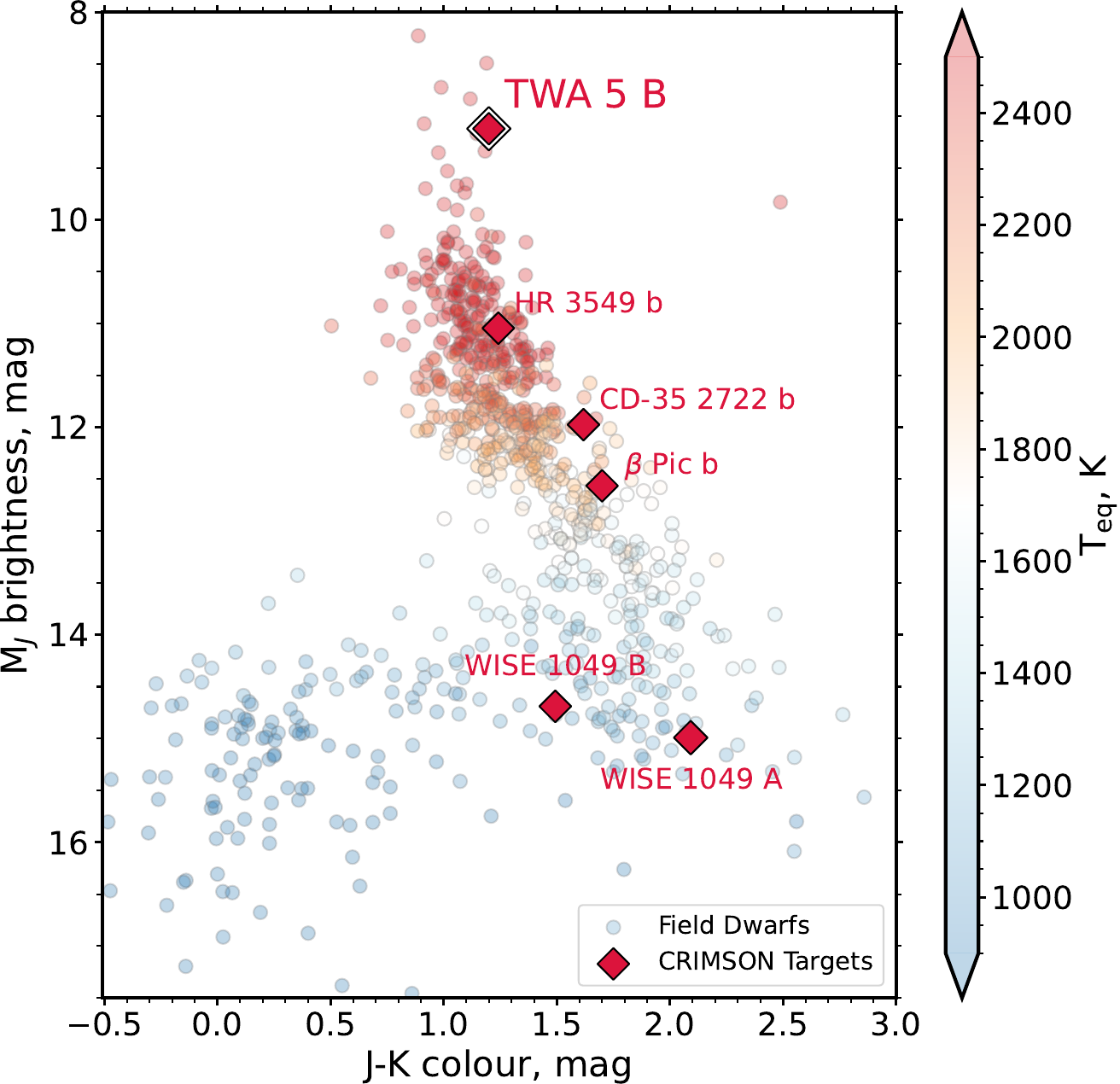}
    \caption{Targets in the CRIMSON survey (Programme ID: 114.27LL.002, PI: Parker), targeting super-Jupiter companions with CRIRES+ \textit{M}-band, plotted on the colour-magnitude diagram of sub-stellar objects. The target in this work, TWA\,5\,B (highlighted), is the hottest target in this survey, at the M/L transition.}
    \label{fig:survey_plot}
\end{figure}

The CRIMSON survey\footnote{CRIres+ \textit{M}-band Spectroscopy of Outer compaNions} (Programme ID: 114.27LL.002, PI: Parker) is designed to trace silicon chemistry in the \textit{M}-band across six planetary mass companions, with targets spanning effective temperatures from 2400 to 1200 K (M8.5 to T0.5). This temperature range covers the onset and then breakup of magnesium silicate clouds in the sub-stellar population \citep{Suarez2022}. We use CRIRES+ in the \textit{M}-band from 3.5--5.2~$\mu$m to probe spectral features from CO, H$_2$O, and with the primary goal of detecting the strong spectral features of gaseous SiO at 4 $\mu$m. We select a sample of six super-Jupiters for the survey (see Figure \ref{fig:survey_plot}), with the target selection driven by the constraints on the achievable signal to noise in the \textit{M}-band with CRIRES+. With the exception of the WISE 1049-5319 AB binary, each selected target orbits a bright host star (see Table \ref{tab:selected_targets}), which is necessary for accurate acquisition, AO performance, spectral extraction, and wavelength calibration of the CRIRES+ spectra. Crucially, the planetary mass companions are young, hot, and nearby such that they are observable above the enhanced thermal background noise at \textit{M}-band wavelengths \citep{Parker2024}.

\begin{table*}
\small
\setlength{\tabcolsep}{4pt}
\centering
\caption{Targets observed in the CRIMSON survey. All targets are observed in the M4368 grating setting with the 0.2$\arcsec$ slit, and use AO, with the exception of the WISE 1049-5319 AB binary, which is not sufficiently bright to act as a natural guide star and is therefore observed in no AO mode. The association BANYAN $\Sigma$ membership probability is shown in parenthesis for each association. Each directly imaged companion has additionally been observed in the \textit{K}-band, see \citet{deRegt2024}.}
\label{tab:selected_targets}
\begin{tabular}{lccccccc}
\hline
On-slit targets & $d$ (pc) & Association (\%) & Age (Myr) & SpType & $m_K$ (mag) & Mass & Exp.\ Time \\
\hline \hline

\begin{tabular}{@{}l}TWA\,5, \\ TWA\,5\,B\end{tabular}
 & $49.6\pm0.1^{(1)}$ 
 & TW Hya (99.9\%)$^{(2)}$ 
 & $10\pm3^{(3)}$ 
 & \begin{tabular}{@{}l}M1.5, \\M8.5$^{(4)}$\end{tabular}
 & \begin{tabular}{@{}l}$7.39\pm0.04$, \\$11.4\pm0.2^{(5)}$\end{tabular}
 & \begin{tabular}{@{}l} $0.9 \pm 0.1$ $M_{\rm \odot}$ $^{(6)}$, \\$25\pm5$ $M_{\rm J}$ $^{(7)}$\end{tabular}
 & 2.2~h \\[12pt]

\begin{tabular}{@{}l}CD-35 2722, \\ CD-35 2722 B\end{tabular}
 & $22.36\pm0.01^{(1)}$ 
 & AB Doradus (99.9\%)$^{(2)}$ 
 & $149^{+51}_{-19}{}^{(3)}$
 & \begin{tabular}{@{}l}M1$^{(8)}$, \\L4$^{(9)}$\end{tabular}
 & \begin{tabular}{@{}l}$7.03\pm0.05$, \\$12.01\pm0.07^{(10)}$\end{tabular}
 & \begin{tabular}{@{}l} $0.4 \pm 0.05$ $M_{\rm \odot}$ $^{(9)}$, \\$31\pm8$ $M_{\rm J}$ $^{(10)}$\end{tabular}
 & 1.7~h \\[12pt]

\begin{tabular}{@{}l}HR\,3549, \\ HR\,3549 B\end{tabular}
 & $94.8\pm0.3^{(1)}$ 
 & Field (92.1\%)$^{(2)}$
 & $125 \pm 25^{(10)}$
 & \begin{tabular}{@{}l}A0V, \\ L0$^{(10)}$ \end{tabular}
 & \begin{tabular}{@{}l}6.04, \\ - \end{tabular}
 & \begin{tabular}{@{}l} $2.375 \pm 0.070$ $M_{\rm \odot}$, \\ $45 \pm 5$ $M_{\rm J}$ $^{(10)}$ \end{tabular} 
 & 0.3~h \\[12pt]

\begin{tabular}{@{}l}$\beta$\,Pic, \\ $\beta$\,Pic b\end{tabular}
 & $19.44\pm0.05^{(1)}$ 
 & $ \beta$ Pictoris (99.9\%)$^{(2)}$ 
 & $24\pm3^{(3)}$
 & \begin{tabular}{@{}l}A5V, \\ L2$^{(11)}$\end{tabular} 
 & \begin{tabular}{@{}l} 3.48, \\ $12.30\pm0.15^{(11)}$\end{tabular} 
 & \begin{tabular}{@{}l} $1.75^{+0.03}_{-0.02}$ $M_{\rm \odot}$, \\ $11.90^{+2.93}_{-3.04}$ $M_{\rm J}$ $^{(12)}$ \end{tabular} 
 & 2.8~h \\[12pt]

\begin{tabular}{@{}l}WISE 1049 A, \\ WISE 1049 B\end{tabular}
 & $2.02\pm0.15^{(13)}$ 
 & Argus (94.1\%)$^{(2)}$ 
 & $40$--$50^{(14)}$
 & \begin{tabular}{@{}l}L7.5, \\T0.5$^{(15)}$\end{tabular}
 & \begin{tabular}{@{}l}$9.44\pm0.07$, \\$9.73\pm0.09^{(15)}$\end{tabular}
 & \begin{tabular}{@{}l}$33.5\pm0.3$ $M_{\rm J}$, \\$28.6\pm0.3$ $M_{\rm J}$ $^{(16)}$ \end{tabular}
 & 0.7~h \\
\hline \hline
\end{tabular}

\par\smallskip
\noindent\textbf{References.} 
    (1)~\citet{Gaia_2020};
    (2)~\citet{Gagne_ea_2018};
    (3)~\citet{Bell_ea_2015};
    (4)~\citet{Lowrance_ea_1999};
    (5)~\citet{Patience_ea_2012};
    (6)~\citet{Kohler2013};
    (7)~\citet{Weintraub2000};
    (8)~\citet{Torres_ea_2006};
    (9)~\citet{Wahhaj_ea_2011};
    (10)~\citet{Mesa_ea_2016};
    (11)~\citet{Chilcote_ea_2017};
    (12)~\citet{Lacour_ea_2021};
    (13)~\citet{Luhman_ea_2013};
    (14)~\citet{Zuckerman_2018};
    (15)~\citet{Burgasser_ea_2013};
    (16)~\citet{Lazorenko_ea_2018}

\end{table*}

Second, each target is chosen such that SiO is potentially detectable based on the predicted abundance from equilibrium chemistry. The highest temperature targets in the sample, TWA\,5\,B and HR 3549 b, have predicted TP profiles from radiative convective thermal equilibrium that are hot enough to avoid silicate cloud formation, ensuring that the atmospheric silicon remains locked in gaseous SiO and is thus a tracer of accreted refractory elements during formation. Further \textit{M}-band CRIRES+ data of $\beta$ Pic b (Programme ID: 13.26UN.001, Co-PIs: Birkby, Parker) is incorporated to follow up the tentative SiO signal observed in \citet{Parker2024}, but not recovered in subsequent data \citep{Janson2025}. CD-35 2722 b is selected as a $\beta$ Pic b analogue to provide a complementary comparison to the presence of SiO in $\beta$ Pic b, in the temperature regime in which SiO abundance is governed by cloud condensation. WISE 1049-5319 AB (Luhman 16 AB) is a binary brown dwarf, but is included in the sample as it has previously been observed to have patchy clouds, and may therefore show SiO emission through gaps in the cloud coverage  \citep{Crossfield2014, Biller2024, Chen2024, Chen2025}.

Finally, each target has archival \textit{K}-band observations from the ongoing CRIRES+ large survey of super-Jupiters (Program ID: 110.23RW, PI: Snellen), which is measuring C/O and $^{12}$C/$^{13}$C for each target (e.g. \citealt{deRegt2024}). Our proposed \textit{M}-band observations are designed to optimise this synergy, and to additionally provide not only the unique access to silicon chemistry, but also independent measurements of the C/O and potentially the $^{12}$C/$^{13}$C ratio, ensuring that these constraints are independent of wavelength dependent effects, e.g. cloud opacities, which have a reduced scattering impact in the \textit{M}-band. The targeted \textit{M}-band detection of SiO will also provide crucial context of the cloud conditions, probed through the depletion of the SiO abundance, which will measure the amount of oxygen sequestered into clouds, which biases the C/O ratios observed in the \textit{K}-band. As a pathfinder instrument for the METIS/ELT LM-band IFU (R$\sim$100\,000; \citealt{Brandl2021}), wide-scale use of CRIRES+ \textit{M}-band is vital preparation for the next generation of instrumentation (see \citealt{Parker2024}).

\subsection{The TWA\,5 system}
\label{sec:TWA}

TWA\,5 A (CD-33 7795) is a pair of pre main sequence T Tauri stars (denoted as TWA\,5 Aa and Ab) with a six year period (6.03~$\pm$~0.01 yrs; \citealt{Mohanty2003, Kohler2013}). When viewed as an unresolved binary TWA\,5 A has an M1.5 spectral type, a total system mass of 0.9~$\pm$~0.1~M$_{\odot}$, and a mass ratio of 1.3~$^{+0.6}_{-0.4}$ between the two components \citep{Kohler2013}. TWA\,5 A is a bright (K = 7.39 mag) member of the TW Hya association, at a distance of 49.6 $\pm$ 0.1 pc \citep{Gaia2020}. TWA\,5 A is not observed to show an infrared excess, with an upper limit on a warm disc luminosity L$_{\text{IR}}$/L$_{\star}$ $\sim$ 7$\times$10$^{-3}$, suggesting there is no significant circumstellar material surrounding the host \citep{Metchev2004, Weinberger2004, Uchida2004}. Therefore, while some members of the TW Hya association maintain their primordial gas discs (most notably the transition disc surrounding the archetypal TW Hydra; \citealt{Andrews2016}), the TWA\,5 system appears to have undergone disc dispersal. The system age is estimated to be 10~$\pm$~3~Myrs through isochrone fitting, corroborated by the observation of H$\alpha$ emission and weak Na I absorption from TWA\,5\,B \citep{Bell_ea_2015,Weintraub2000, Neuhauser2000}.

The sub-stellar companion TWA\,5\,B orbits the stellar binary, and is located at a separation of 1.8" ($\sim$86 AU; \citealt{Neuhauser2010}). TWA\,5\,B was first suggested as an imaged companion candidate \citet{Lowrance1999, Webb1999} and confirmed by proper motion and spectra by \citet{Neuhauser2000}. Low-resolution spectra from FORS2 (visible), ISAAC (\textit{H}-band), SINFONI (\textit{JHK}-bands), Magellen-AO/Clio2 (\textit{L}-band) and HST/STIS (UV) suggest an M8.5 spectral type, a 25~$\pm$~5~M$_J$ mass, $\log(g)$ = 3.5--4.0, and T$_{\text{eq}}$ $\approx$ 2400 K \citep{Neuhauser2000,Schneider2000, Weintraub2000, Bonnefoy2014, Stone2016}. The spectral types of brown dwarfs and directly imaged planets do not conform to the standard Morgan-Keenan classification system and young, hot, planetary mass objects therefore have spectra resembling M dwarf stars of the same photospheric temperatures, despite their distinctly sub-stellar masses. TWA\,5\,B has also been detected at X-ray wavelengths \citep{Tsuboi2003}, demonstrating a very soft spectrum (i.e. a spectrum composed of lower energy X-rays produced by thermal emission) with a dominant plasma temperature of 0.3 keV, implying hot coronal gas displaying magnetic activity. In this work we follow the naming convention in the literature, and denote the sub-stellar companion as a brown dwarf, TWA\,5\,B. We note, however, that the atmospheric properties of TWA\,5\,B are analogous to those of hot directly imaged planets, and both our analysis techniques and discussion centre on the application of these techniques to exoplanetary objects. In this work we will refer to the unresolved stellar binary TWA\,5 Aab as `the host' and the bound sub-stellar companion TWA\,5\,B as the `companion'.

\section{Observations}
\label{sec:Observations}

We observe the TWA\,5 system with the Cryogenic High-Resolution In-frared Echelle Spectrograph \citep[CRIRES+;][]{Dorn2014,Dorn2023}, mounted at the Nasmyth B focus of the VLT UT3 (Melipal). Our observations (see Table \ref{tab:obs_table}) are taken on the night of 19/03/2025 UTC and consist of two 1.6 hour observing blocks (OBs), which were executed consecutively (Program ID: 114.27LL, PI: Parker). CRIRES+ operates as a cross-dispersed slit spectrograph, with a 10\arcsec long slit, which we orient with a position angle PA~=~350.25\degr such that both the host star and the target companion TWA\,5\,B are aligned along the slit. We select the slit width of 0.2\arcsec to achieve the maximum spectral resolving power offered by CRIRES+ in the \textit{M}-band, nominally R~$\sim$~92~000 for observations with uniform illumination of the slit, while minimising the inclusion of excess sky background noise. Each of the two observing blocks consists of 200 science exposures with an exposure time of 20s. The observations are taken in an ABBA nod pattern during which the telescope is nodded 4.5\arcsec along the slit, to allow accurate background subtraction which is crucial in the background-dominated \textit{M}-band. To minimise overheads we observe five exposures at each nodding position, but do not stack exposures before nodding, resulting in a sequence of five A exposures, followed by ten B exposures, then five A exposures for each complete nodding cycle (NDIT=1, NEXP=5). In the \textit{LM}-bands the sky background remains stable over timescales $<$ 150s, and this strategy maintains an acceptable duty cycle while permitting sufficiently precise background subtraction. We observe using the M4368 grating setting, which provides non-continuous wavelength coverage across the range 3.51--5.21~$\mu$m over six orders (see \citealt{Parker2024}). The third spectral order lies in a region of total atmospheric opacity due to the 4.4~$\mu$m telluric CO$_2$ band and is excluded from our analysis for both OBs. 

\begin{figure}
    \includegraphics[width=\columnwidth]{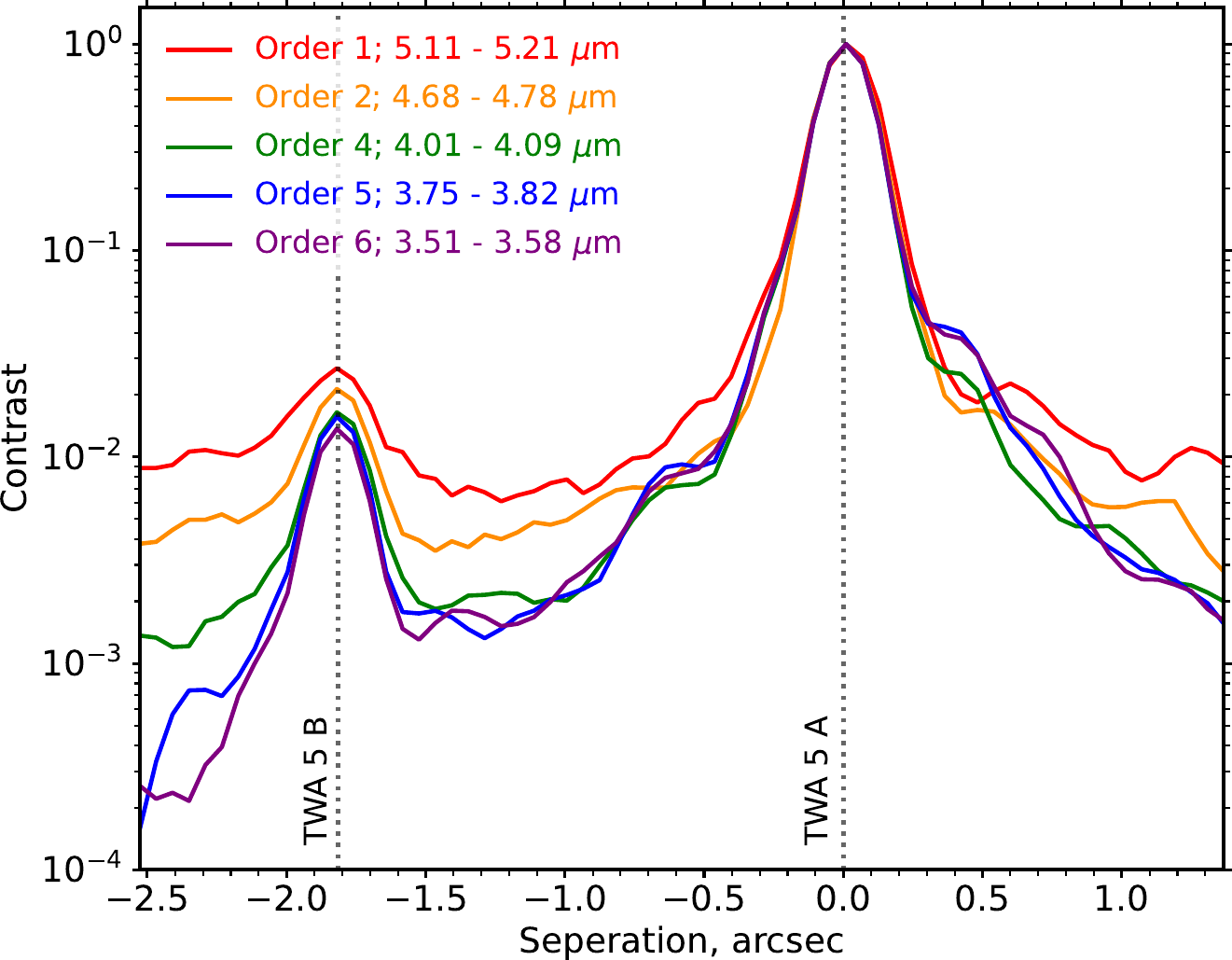}
    \caption{Spatial profile of our observations, demonstrating the contrast achieved for each spectral order following summing the spectral dimension across the full time series. The companion TWA\,5\,B is clearly visible in the photometric contrast, but the companion spectrum is heavily contaminated by the telluric contamination and thermal background in the \textit{M}-band.
    }
    \label{fig:contrast}
\end{figure}

The observations were taken at low airmass ($<$ 1.2) in good observing conditions, with average DIMM measured seeing of 0.52\arcsec and 0.66\arcsec, and radiometer measured PWV of 1.2, 1.6 mm for OB1 and OB2, respectively. The use of the Multi-Application Curvature Adaptive Optics system (MACAO; \citealt{Paufique2004}), with TWA\,5\,A as the AO natural guide star, results in a suppression of the starlight of the order of 10$^{-3}$ at the position of the substellar companion (see Figure \ref{fig:contrast}), separated by 1.8$\arcsec$ from the host star. Following each observing block we additionally observe a telluric calibration star in close spatial proximity to the target (11\arcmin), to assist with the telluric correction of the science spectra (see Section \ref{sec:post_proc}). For our telluric standard we observe the bright B9V star Omicron Hydrae for ten exposures following each OB, with the same observing strategy as used for the science frames (T$_{\text{exp}}$ = 20, NDIT=1, NEXP=5), for a total of 0.2 h including overheads.

\begin{table}
	\centering
	\caption{Details of our CRIRES+ observations of TWA\,5\,B. The average precipitable water vapour (PWV) and DIMM measured seeing are recorded from conditions at Paranal at the time of observation.\label{obs_table}}
	\label{tab:obs_table}
	\begin{tabular}{lcc} 
		\hline
             & OB1 & OB2\\
		\hline
            \hline
            Date & 19/03/2025 & 19/03/2025 \\
            Av.~PWV & 1.19~mm & 1.61~mm \\
            Av.~Seeing & 0.524\arcsec & 0.659\arcsec \\
            Av.~Airmass & 1.02 & 1.11 \\
            N$_{\text{exp}}$ & 200 & 200 \\
            DIT & 20~s & 20~s \\
            Slit Width & 0.2\arcsec & 0.2\arcsec\\
            Position Angle & 350.25\degr & 350.25\degr\\
            Companion Sep. & 1.815\arcsec & 1.815\arcsec\\
            Grating Setting & M4368 & M4368\\
            Resolution & $\approx$~92~000 & $\approx$~92~000\\
            $\lambda$ Coverage & 3.51--5.21~$\mu$m& 3.51--5.21~$\mu$m\\

		\hline
  \hline
	\end{tabular}
\end{table}

\section{Methods}
\label{sec:Methods}

In this work we build upon the analysis methods for \textit{M}-band CRIRES+ data developed in \citet{Parker2024}. Our analysis approach is to extract a spectrum from the detector at each position along the slit (Section \ref{sec:spectral_extraction}) and use these to clean the data of telluric and stellar contributions (Section \ref{sec:post_proc}). We then aim to detect the companion spectral features through cross-correlation with model atmospheric spectra (Section \ref{sec:cross_corr_setup}), and retrieve the atmospheric properties of the companion TWA\,5\,B (Section \ref{sec:models_and_retreivals}).

\subsection{Basic calibrations}

We use \textsc{pycrires}\footnote{\url{https://pypi.org/project/pycrires/}} \citep{pycrires, Landman2024} to perform basic calibrations. The raw frames are flat-fielded and dark subtracted to remove detector and readout artefacts present in the raw exposures. We additionally correct for non-linearity effects at the pixel level using CRIRES+ calibrations, mask bad pixels, and correct for the imprinted instrumental blaze function. To mitigate the sky thermal background present across all \textit{M}-band orders each exposure (A or B) is subsequently subtracted from the closest exposure in the alternative nodding position (B or A) in the observing sequence, providing an effective background subtraction. A robust measurement of slit curvature is crucial to the success of spatially resolved spectroscopy in the \textit{M}-band, since the projection of static spectral features will shift across the spatial extent of each order \citep{Parker2024, Janson2025}. While this effect is well characterised for CRIRES+ \textit{YJHK} grating settings and can be accurately calibrated using the Fabry Perot Etalon (FPE) system, it remains uncharacterised in the L\textit{M}-bands. Here we adopt the approach used in \citet{Parker2024}, and trace the tilt of the sky emission lines in the raw science frames, providing a measure of the slit tilt for each spectral order. Inaccuracies in the slit tilt modelling propagate to increased uncertainties in the measured companion orbital and rotational velocity, but do not impact the significance of molecular detections.

\subsection{Spectral extraction}
\label{sec:spectral_extraction}

Following the application of basic calibrations to the raw data, we aim to extract a spectrum from every spatial position along the slit at both the A and B nod positions. We extract a 4\arcsec wide region of the slit around the companion and host star, at the 0.056~\arcsec/pix spatial pixel scale of the CRIRES+ detectors, choosing a region with effective background subtraction and which avoids contamination from the negative nodding image. The spectrum from each position along the slit is projected onto the detector following a curved spectral trace which is unique to the detector, spectral order, and grating setting chosen. Every spatial position along the slit contains valuable information; i) the target host star provides a measurement of the contaminating stellar and telluric spectra; ii) each of the spectra at increasing separations from the star contain the PSF shape and measurements of the underlying noise, required to clean the data; and iii) the spatial positions along the slit at the companion separation additionally contain the spectra of the companion TWA\,5\,B. 

\subsubsection{The periodic systematic in 2D CRIRES+ spectra}

The echelle spectra from each order follow a curved trace on the detector for each position along the slit, and must be extracted and rectified onto a Cartesian grid to clean and analyse the data. When carrying out extraction at each spatial position along the slit, or rectifying the detector image onto a Cartesian grid through interpolation, a periodic systematic appears in the extracted spectra, which has previously been identified in \citet{Landman2024} and \citet{Parker2024}. Here, we determine that this systematic arises due to the discrete spatial sampling of the CRIRES+ detector (0.056$\arcsec$/pixel), and that the periodicity of the oscillation is directly linked to the gradient of the spectral trace that is dispersed on the detector. 

\begin{figure*}
    \includegraphics[width=2\columnwidth]{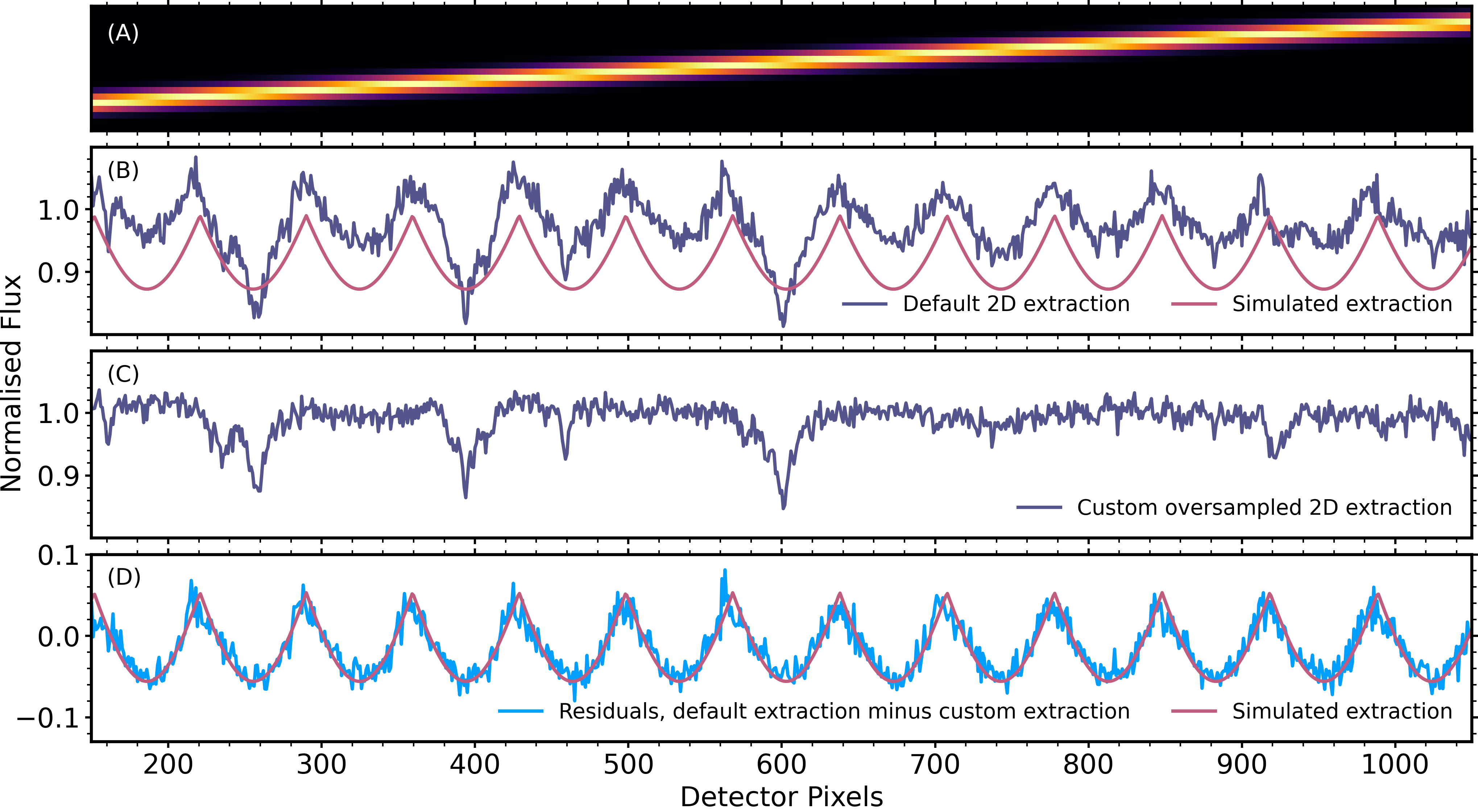}
    \caption{A) The simulated trace of a flat spectrum across the CRIRES+ detector, demonstrating the periodic variations in pixel-level flux as the spectral trace crosses pixels in the spatial (cross-dispersion) direction of the detector. B) The extracted stellar spectrum for an exposure using default boxcar extraction along a 1 pixel wide aperture (purple), which shows clear periodicity that dominates the continuum. This periodicity is matched by the simulated extraction of a flat spectrum (red) which shows how the pixel-level sampling induces periodic variations in the extracted flux as the spectral trace crosses pixels in the spatial direction of the detector. C) The extracted spectrum obtained when using the custom extraction routine proposed in this work and described in Section \ref{sec:Methods}. D) The residuals between the default extraction method and the custom extraction method (blue), compared to the simulated extraction of a flat spectrum. We demonstrate that the phase, shape, and magnitude of the extraction systematic is accounted for by the detector-level effects, and the use of the custom oversampling extraction routine can mitigate this periodicity.}
    \label{fig:systematic_plot}
\end{figure*}

To demonstrate that this systematic arises exclusively due to the pixel sampling of the detector we simulate the extraction of spectra from idealised simulations of the CRIRES+ detectors. Taking the known locations of the spectral traces measured from the M4368 science frames, we inject a flat spectrum (i.e. a spectrum of ones) following these traces onto an empty 2048$\times$2048 grid, with the x-axis the spectral pixels on the detector (the dispersion direction), and the y-axis containing the spatial pixels along the slit (the cross-dispersion direction). The injected flat spectrum has a spatial distribution of a Gaussian with height one and a width matching the measured PSF width (Figure \ref{fig:systematic_plot} panel A). From this simulated detector frame we then extract spectra at each spatial pixel following both the default extraction routines for spectral extraction, and the proposed custom oversampling method presented in this work. For the default extraction routine the extracted stellar spectrum shows a clear periodicity that dominates the continuum, of the order of $10\%$ of the extracted flux, shown in red in Figure \ref{fig:systematic_plot} panel B. This periodicity directly matches the phase and magnitude of the periodicity seen in the extracted science frames when using default extraction (see Figure \ref{fig:systematic_plot} panel B). The origin of this periodicity is determined to be the discrete sampling of the PSF in the spatial direction. When the PSF is sampled directly at its peak, with the spectral trace lying directly at the centre of a spatial pixel, the extracted flux is at a maximum. In comparison, if the spectral trace has its peak between two spatial pixels, the extracted flux is lower, as the full peak of the PSF is not reconstructed in the default extraction. This leads to the observed periodicity, which is present in all CRIRES+ frames in all grating settings, but only becomes a challenge when spectra must be individually extracted at each spatial position. For the typical science cases in which a single 1D spectrum of the target is extracted, for example observations of short period planets, the periodicity is mitigated by extracting over many spatial pixels, and by oversampling the PSF. However, when the spatial information must be preserved (e.g. for directly imaged systems) we must extract a spectrum at each spatial pixel. Additionally, in the case of very high thermal background, extraction over a larger number of pixels significantly increases the included background noise. We note that the period of the systematic is highly dependent on the angle of the trace on the detector, and therefore grating settings with a low dispersion angle (e.g. K2166; \citealt{Landman2024}) see a reduced impact compared to those with a steeper angle (e.g. M4368; \citealt{Parker2024}).

\subsubsection{An extraction routine to mitigate the sampling systematic}

To mitigate this systematic effect, we develop an oversampling extraction routine. First, all frames for each nod position and observing block are stacked, and the location of the stellar spectrum on the detector is traced using a 2\textsuperscript{nd} order polynomial, providing a measure of the spectral trace. The data is subsequently oversampled by a factor of 100 in the spatial direction using 4\textsuperscript{th} order spline interpolation to reconstruct the PSF, after which the spectral trace is used to extract a spectrum at each spatial position along the slit, using a three spatial pixel wide boxcar extraction (0.168$\arcsec$ wide; 300 spatial pixels of the oversampled image). The three pixel width of the boxcar extraction is chosen to maximise the included signal while minimising the impact of the systematic effect, and suppress the inclusion of excess background noise. This extraction routine suppresses the systematic to $\leq$ 10$^{-3}$ times the extracted flux level, and thus this effect is sub-dominant to the thermal background noise, which is the primary noise source in the \textit{M}-band exposures. This suppression is achieved through two approaches. First, by spatially oversampling the spectra, we can reconstruct the PSF for each spectral pixel, ensuring that the sampling of the PSF is constant at all spectral pixels, thus removing the periodicity. Second, by extracting across a three spatial pixel wide aperture, we can maintain spatial information, but average over any remaining periodicity. Finally, we highlight that this effect should be common to all cross-dispersed echelle spectrographs if used for science cases that require spatial information to be extracted, including METIS-IFU.

\subsection{Subsequent processing}

Following extraction of the spectra we correct for the impact of the slit tilt by rectifying the spectral offset of each extracted spectrum from the central stellar spectrum using flux conserving interpolation, and following the measured slit tilt. We subsequently perform wavelength calibration of the spectra, collapsing the spatial information of each exposure into a single 1D spectrum to provide a higher S/N master spectrum with which to compare to a stable wavelength reference. Due to the absence of suitable stable wavelength reference sources in CRIRES+ in the \textit{M}-band, we use a telluric model from ESO SkyCalc \citep{Noll2012, Jones2013} as the stable reference source. We adopt a 2\textsuperscript{nd} order polynomial as the M4368 wavelength solution and generate a grid of wavelength solutions, varying each of the parameters in the 2\textsuperscript{nd} order fit. To perform wavelength calibration we iteratively cross-correlate a telluric model with each wavelength solution with the host star flux for each exposure. The model which maximises the cross-correlation for each order is considered the best match to the wavelength solution and adopted as the prior for the next grid of models, with this process iterating with a refining grid size until a stable solution is reached. We estimate an average precision of the order of $\pm$ 5$\times$10$^{-3}$ nm which is measured from the FWHM of the cross-correlation functions, corresponding to $\pm$ 0.17 km/s precision.  

Any remaining bad pixels or outliers on the scale of a single pixel that remain in the extracted spectra are detected through application of a Laplacian of the Gaussian algorithm (e.g. \citealt{Kong2013}, see \citealt{vanSluijs2023}). Pixels that deviate by $>$~5$\sigma$ are interpolated over using the nearest pixels in the dispersion direction. We note that the asymmetry of the CRIRES+ M4368 PSF (see the shoulder of the PSF at 0.5\arcsec in Figure \ref{fig:contrast}; seen also as the `secondary trace' in \citealt{Parker2024} observations of $\beta$ Pic) is confirmed to be an effect that is independent of the source observed, and is seen in all observations of TWA\,5 and calibration stars with the M4368 grating setting.


\subsection{Post-processing}
\label{sec:post_proc}
The companion spectrum, which is visible when stacking all observed frames, is nonetheless obfuscated by three major contaminants: the Earth's telluric spectrum, the spectra from the central stellar binary, and the thermal background noise. The post processing of the data attempts to isolate and remove each contaminant in turn, leaving the companion spectral features buried in a noise pattern that is no longer spatially or temporally correlated, from which they can be extracted using cross-correlation with appropriate atmospheric models. Our post processing procedure is as follows. First, we generate models of the telluric absorption spectrum using \textsc{molecfit} \citep{Smette2015} to fit the observed telluric calibration star spectra. Due to the plethora of absorbing molecules in the 3.51 -- 5.21 $\mu$m region, we fit abundances of the telluric absorbers H$_2$O, CH$_4$, N$_2$O, CO, O$_3$, CO$_2$, and OCS. We select limited wavelength regions within each order with high S/N telluric lines to perform the fitting, avoiding regions of saturated tellurics, and adopt a 2\textsuperscript{nd} order continuum fit. The stellar spectrum in the science frames cannot be used for generating a telluric model with \textsc{molecfit}, due to the near-total coverage of M-dwarf spectral lines \citep{Parker2025}. The largest variations in the telluric contamination of our science frames during our observations are driven by the differential in the airmass of the target, and from PWV variations on timescales of minutes. Consequently, we recalculate the modelled \textsc{molecfit} telluric transmission spectrum for each science exposure, using the fixed global chemical abundances of molecular species derived from the \textsc{molecfit} fitting to the standard star, but scaling the airmass following the Beer-Lambert law, and the H$_2$O abundance using the radiometer measurements of PWV at the epoch of each exposure. 

The science spectra are divided through by the model telluric spectra, and regions of the model telluric spectra with $\leq$ 70$\%$ transmission are masked to remove the high error regions in the telluric cores. We additionally mask any high variance spectral channels that exceed three times the median standard deviation, representing unreliable telluric fits to the data. Ten columns at both edges of each order which show enhanced systematics are additionally masked. This use of \textsc{molecfit} in the \textit{M}-band represents a significantly increased challenge when compared to the routine use on optical high-resolution data (e.g. \citealt{Langeveld2021, Maguire2024}).

Following the telluric removal process, the stellar spectrum dominates the flux in the data, but is localised in the stellar PSF. The stellar binary is classified as an M1.5 spec type, and contains features from H$_2$O and CO in its photosphere. Unlike in the case of \citet{Parker2024}, we choose not to model and remove the stellar component due to the large separation between the sub-stellar companion and stellar host and instead employ an implementation of principal component analysis (PCA) to capture and remove the stellar spectrum without eroding the companion signal. To do this we stack all exposures, and then apply PCA to our data cube in the wavelength domain (e.g. \citealt{deKok2013, Damiano2019, Ridden-Harper2016}), as opposed to the usual application of PCA in the spatial domain for directly imaged targets (e.g. \citealt{Landman2024, Parker2024}). This choice is motivated by the fact that wavelength-domain PCA enables us to robustly exclude the spatial separations containing the companion signal from the calculation of the eigenvectors. We can therefore build a systematics model for all spatial positions that is not informed by the spectra at the companion position, and does not erode the signal from the companion. As an M8.5 object, the spectrum of TWA\,5\,B shares features of H$_2$O and CO with the M1.5 host star spectrum, and the companion signal is sufficiently strong that it is heavily eroded if the position of TWA\,5\,B is not excluded from the calculation of the eigenvectors. Therefore, we exclude the data from the five spatial rows which contain the signal of TWA\,5\,B, and then calculate the usual singular value vector decomposition:

\[
A^\prime = U S V^{\top}
\]

where $A^\prime$ is the reconstructed model of the data, $U$ the left singular vectors with orthonormal columns in wavelength, S the diagonal matrix, and $V$ the eigenspectra. The eigenspectra capture the global spectral trends in the data (with the companion position excluded), and these are scaled for each spatial position by the spatial trends in the data, $U$ which contain the companion rows. The reconstructed model $A^\prime$ can thus capture the wavelength dependent systematics which dominate the datacube from the stellar spectrum, telluric residuals, and thermal background, while minimally impacting the signal from TWA\,5\,B. We subsequently remove the systematics model reconstructed using the chosen PCA components, adopting a standard deviation based stopping criteria \citep[see][]{Parker2025, Vaughan2026}. We choose a 5$\%$ change in the standard deviation from the previous iteration as the threshold to determine the number of PCA components to remove.

\subsection{Cross-correlation setup}
\label{sec:cross_corr_setup}
\begin{figure*}
    \includegraphics[width=2\columnwidth]{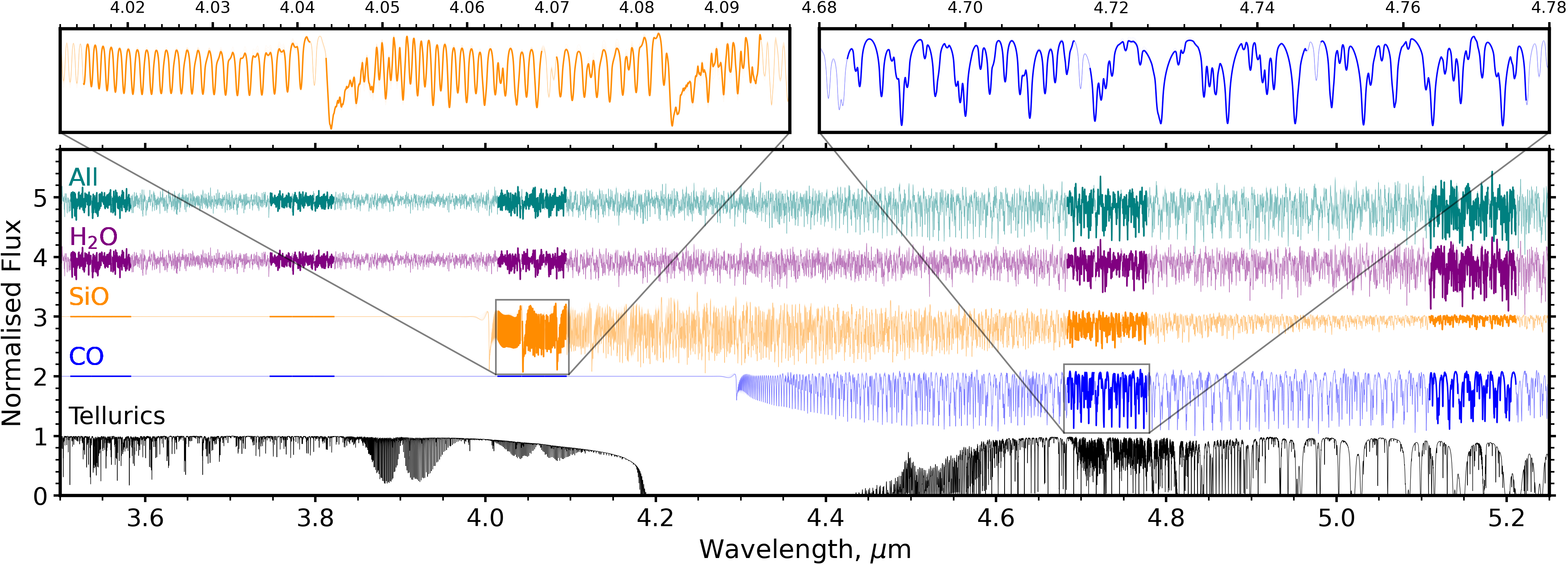}
    \caption{Best fitting atmospheric model from the retrieval analysis containing all species (teal), and corresponding models with the contribution from individual molecular absorbers (H$_2$O, SiO, CO) only. For visual clarity these models have been continuum normalised by convolution with a Gaussian kernel and rescaled based on the maximum depth of their spectral lines, and therefore the relative depths of the spectral lines between models are for illustrative purposes only. These models are used to cross-correlate with the data to produce the molecular detections in Figure \ref{fig:4xdet}. The highlighted regions of the models indicate the wavelengths observed by CRIRES+ in the M4368 grating setting, with the inset axes demonstrating the regular structure of the dominant band heads of SiO and CO which drive the detections of these molecules. The model telluric spectrum (black) demonstrates the extreme contamination at these wavelengths.}
    \label{fig:models}
\end{figure*}

Following the removal of telluric contamination using \textsc{molecfit}, and the subsequent application of PCA, the residual spectra at each slit position for each exposure are primarily composed of noise. The exception is the spatial position containing the companion, which includes an additional contribution from the spectrum of TWA\,5\,B. To detect molecular features in the spectrum of TWA\,5\,B we stack all observed frames, then cross-correlate with template atmospheric spectra (Figure \ref{fig:models}), generated through radiative transfer modelling, see Section \ref{sec:models_and_retreivals}. We calculate the Pearson correlation coefficient for each spatial position, across velocity shifts of $\pm$ 300 km/s, sampled at the CRIRES+ resolution element width of 1.5 km/s (see Section \ref{sec:Discussion}; Figure \ref{fig:4xdet}). The S/N is calculated following \citet{Parker2024}, by dividing the CCF by the standard deviation of each row, excluding the companion orbital velocities. To account for correlated noise structures in the CCF induced by the rotational broadening of the spectral lines, we calculate the standard deviation of each row from the downsampled CCF, where the CCF is sampled at every Nth point, with N tuned to $\sqrt{2}$ $v_{\text{rot}}$. Here, $v_{\text{rot}}$ is the measured rotational velocity of the companion signal (14.19$^{+0.88}_{-0.79}$ km/s) and represents the region over which we expect signals to be correlated (see \citealt{Parker2024}).

\subsection{Atmospheric modelling and retrieval setup}
\label{sec:models_and_retreivals}

We additionally perform an atmospheric retrieval analysis on the spectrum of TWA\,5\,B to infer its chemical abundances, atmospheric dynamics, and the temperature profile of its atmosphere. To prepare a 1D input spectrum for the retrieval, we take the stacked 2D spectra of all exposures following the cleaning process, and extract the spectrum at the companion position. This extraction is performed by using a weighted sum of the signal from the three spatial pixels containing the majority of the companion spectrum, using the measured PSF of the star as weights for the relative flux contribution of each spatial pixel. 

\subsubsection{Model reprocessing}

When carrying out atmospheric retrievals with high-resolution spectra it is important to apply robust model reprocessing to replicate the impact of the PCA detrending processes on the models used in the retrieval process \citep{Brogi2019, Dash2024}. Although we exclude the spatial rows that contain the companion signal in the calculation of the PCA eigenvectors, the companion spectrum can still be altered by the detrending process. Firstly, the flux from the companion, which is localised at specific spatial positions, can bias the calculation of the spatial components ($U$), which scale the spectral trends captured by the eigenvectors. Second, any minimal leakage of the PSF of TWA\,5\,B beyond the five excluded spatial rows can result in the inclusion of the companion flux in the spectral trends of the systematics model, resulting in the distortion of the targeted atmospheric spectral features. The five excluded spatial rows are chosen as a balance between preserving the companion spectrum from erosion, while allowing the PCA to successfully identify and remove the systematics at these spatial positions.

Our model reprocessing routine for each generated model is as follows. First, we take the 2D stacked spectra for each order and detector, prior to the application of PCA detrending. The spectral model to be reprocessed is injected into this matrix at the companion position, with a PSF shape inferred from the measured stellar PSF. The PCA coefficients are recalculated for this new matrix and the systematics model is removed, leaving detrended data which includes the injected and detrended model. We then subtract the original cleaned data from the newly detrended data (which also contains the model injection), leaving the reprocessed model. This method successfully applies the effects of the PCA on the companion spectral features to the models used in the retrieval, and we observe a flattening of the model continuum and small changes in the line depth under the application of increasing PCA components. This reprocessing is carried out on each model generated in the retrieval, and the reprocessed models are subsequently used for the likelihood evaluation (Section \ref{sec:HyDRA}).

\subsubsection{Hydra atmospheric retrievals}
\label{sec:HyDRA}

To perform the atmospheric retrievals we adopt the HyDRA retrieval framework (e.g. \citealt{Gandhi2019,Gandhi2022}), using the GENESIS radiative transfer model and multi-nested sampling for the Bayesian parameter estimation. This framework has previously been adapted to retrieve the atmospheric properties of non-irradiated planets \citep{Gandhi2023}, and demonstrated on CRIRES+ data in the \textit{K}-band \citep{Gandhi2025}. Here, we adapt the retrieval framework by adding cross-correlation to log-likelihood mapping and model reprocessing for directly imaged planets. Following \citep{Brogi2019}, we express the likelihood function as

\begin{equation}
\text{log}(L) = -\frac{N}{2} \text{log}[s_f^2 -2R(s) + s_g^2],
\end{equation}

in which $s_f^2$ is the variance of the data, $s_g^2$ the variance of the model, and $R(S)$ the cross-covariance

\begin{equation}
s_f^2 = \frac{1}{N} \sum_{n} f^2(n)
\end{equation}

\begin{equation}
s_g^2 = \frac{1}{N} \sum_{n} g^2(n-s)
\end{equation}

\begin{equation}
R(S) = \frac{1}{N} \sum_{n} f(n)g(n-s).
\end{equation}

$f(n)$ represents the mean-subtracted values of a row from each order of the processed data, and $g(n)$ the corresponding processed model template, where $n$ is an individual spectral channel. $s$ denotes a wavelength shift, and $N$ is the total number of spectral pixels used in the calculation. We adopt a free chemistry approach, letting the abundance of each molecule to vary freely, which provides better flexibility in testing silicon abundance and volatile-to-refractory ratios beyond an equilibrium chemistry retrieval. Our atmospheric models contain opacities from the three detected atmospheric constituents; H$_2$O ($^{1}$H$_2$$^{16}$O; \citealt{Polyansky2018, Tennyson2024}), CO ($^{12}$C$^{16}$O; \citealt{Rothman2010, Li2015}), and SiO \citep[$^{28}$Si$^{16}$O;][]{Yurchenko2022}. Each spectral line is broadened using a Voigt profile (e.g. \citealt{Gandhi2017}) at each pressure and temperature grid point using pressure broadening coefficients for H$_2$ and He \citep{Gandhi2020_linelists}. The overall cross-section from each species is subsequently produced by summing this contribution for each spectral line. We adopt abundance profiles of H$_2$O, CO and SiO with constant abundances at all heights, and retrieve their volume mixing ratio (VMR) as a free parameter, following \citet{deRegt2024, Gandhi2025}. We additionally include opacity from collisionally induced absorption of H$_2$–H$_2$ and H$_2$–He \citep{Richard2012}. Alongside gas phase opacity, we include a cloud deck parametrisation, following \citet{Molliere2020} and \citet{Gandhi2025}. We adopt a cloud opacity parametrised by the following free parameters; a grey opacity, $\kappa$, a cloud deck pressure, P$_{\text{cl}}$, and a power law in pressure with the exponent $\gamma_{\text{cl}}$. We thus define the cloud opacity $\kappa_{cl}$ at each pressure layer P for all wavelengths as:

\begin{equation}
\kappa_{\mathrm{cl}}(P) =
\begin{cases}
\kappa \left( \dfrac{P}{P_{\mathrm{cl}}}\right)^{\gamma_{\mathrm{cl}}} & P \le P_{\mathrm{cl}}, \\
0 & P > P_{\mathrm{cl}}.
\end{cases}
\end{equation}

Spectra are generated at a resolution of R = 300\,000, and convolved to the instrumental resolution of R = 92\,000 with a Gaussian kernel, followed by rotational broadening by a kernel corresponding to the vsin(i) of TWA\,5\,B. For the temperature-pressure (TP) profile, we adopt a flexible four-layer parametrisation (e.g. \citealt{Zhang2023, Kothari2024}). The temperature-pressure relation at each point follows a negative power law, with a dimensional exponent $\alpha$ that varies with altitude, defined as 

\[
\alpha = \frac{-d \ln(T)}{d \ln(P)}.
\]

Note the negative gradient, such that a positive $\alpha$ corresponds to a temperature inversion. The exponent $\alpha$ is defined at and interpolated through five specific pressure knots (top of atmosphere, bottom of atmosphere, and three intermediate points), with linear interpolation between each knot. The five pressures for the knots are:
\[
P_g = \left[ P_{\max},\; P_b,\; P_{\mathrm{set}},\; P_t,\; P_{\min} \right],
\]
with corresponding $\alpha$ values:
\[
\alpha_g = [\alpha_0, \alpha_1, \alpha_2, \alpha_3, \alpha_4],
\]

where the pressure knots $P_b$ and $P_t$, are defined by their deviation from the central knot ($T_{\mathrm{set}}$, $P_{\mathrm{set}}$):
\[
P_b = P_{\mathrm{set}} + dP_{\mathrm{set},b},\\
P_t = P_{\mathrm{set}} - dP_{\mathrm{set},t},
\]

and the TP profile is anchored to the central knot ($T_{\mathrm{set}}$, $P_{\mathrm{set}}$). Due to the filtering of the continuum in the reduction process, required to detrend the strong telluric contamination in the \textit{M}-band, these observations have limited constraining power on the companion radius (R$_{\text{p}}$), or surface gravity $\log(g)$. These values are therefore predominately constrained by the adoption of Gaussian priors, following established approaches for high spectral resolution data of directly imaged targets \citep{Landman2024}. We adopt the values and uncertainties of R$_{\text{p}}$ = 2.7 $\pm$ 0.8 R$_{\text{J}}$ and $\log(g)$ = 3.9 $\pm$ 0.2 for the Gaussian priors, based on the 2$\sigma$ uncertainties from previous observations of TWA\,5\,B \citep{Weintraub2000, Patience2012}. We adopt uniform priors for all other retrieval parameters, see Table \ref{tab:combined_table}. We carry out nested sampling using {\tt MultiNest} \citep{feroz_multinest_2009} implemented as {\tt pymultinest} \citep{buchner_x-ray_2014}, running it in constant efficiency mode with a constant sampling efficiency of 0.05 and 5000 live points.

\section{Results}
\label{sec:Results}

\begin{figure*}
    \includegraphics[width=2\columnwidth]{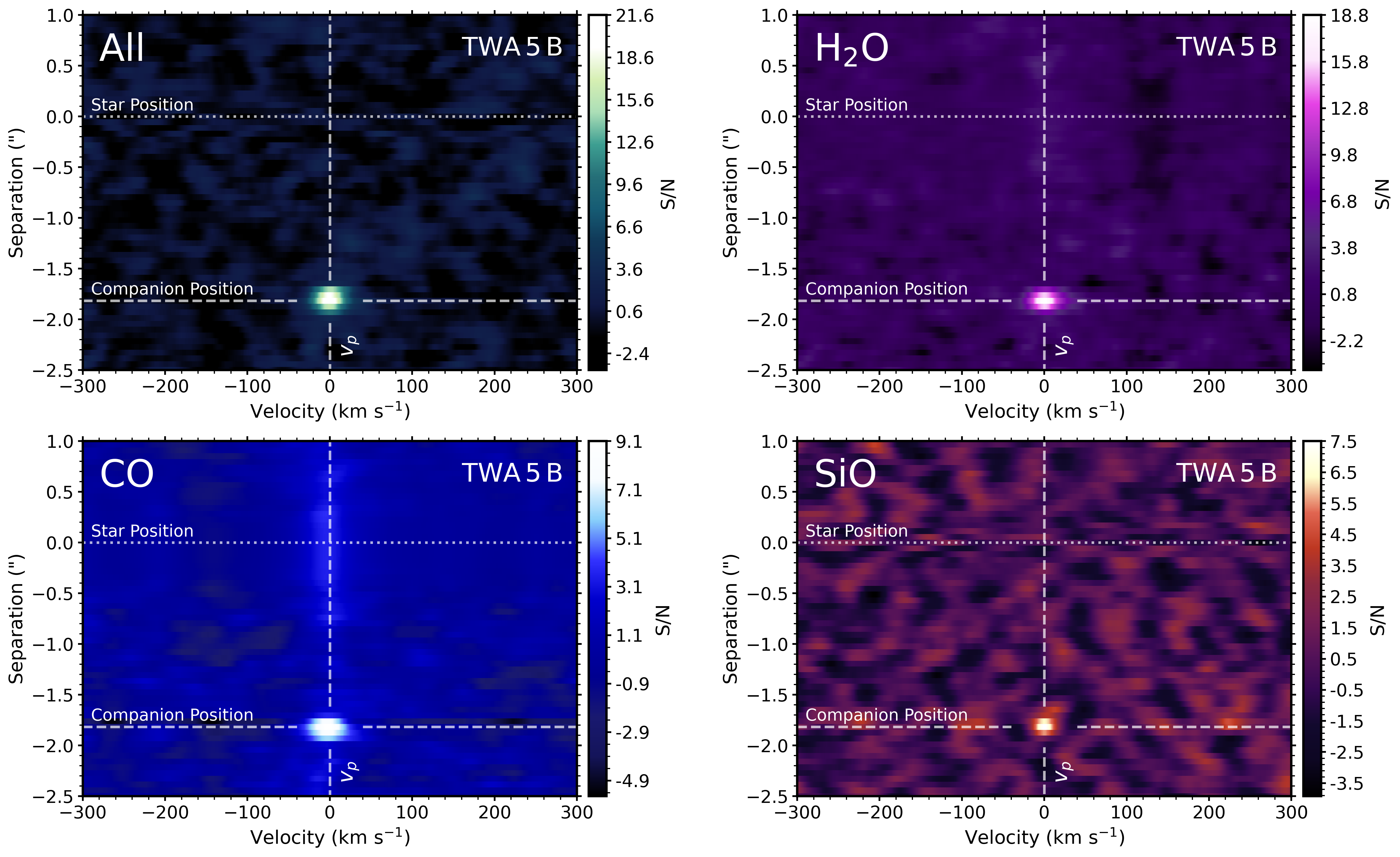}
    \caption{Cross-correlation maps showing the detections of molecular species in the atmosphere of TWA\,5\,B, in the rest frame of the companion. The cross-correlation function (CCF) shows a peak at the velocity (v$_{\text{p}}$) and separation of the companion for each detected molecule. These CCFs are produced using the best fitting models inferred from the atmospheric retrieval (see Section \ref{sec:models_and_retreivals}). The CCF for CO demonstrates contamination from the residual M-dwarf CO lines at separations of $\pm0.5\arcsec$, but this contamination is suppressed at the wide separation of the companion. The velocity width of the cross-correlation peaks for each species is determined by the intrinsic shape of the spectral lines, as explored in \citet{Parker2024}, while the observed variation in the spatial width (y-axis) of the cross-correlation peaks arises from the relative strengths of the signal against the background noise.}
    \label{fig:4xdet}
\end{figure*}

\subsection{Cross-correlation detections}

Cross-correlating with atmospheric model templates, we observe detections of CO (S/N = 9.1), H$_2$O (S/N = 18.8), and SiO (S/N = 7.5) in TWA\,5\,B (Figure \ref{fig:4xdet}). The cross-correlation for each molecule peaks at the predicted separation ($\approx$ 1.8$\arcsec$) and orbital velocity of the companion. Spectral features from H$_2$O dominate the spectrum of the TWA\,5\,B, pervading across all spectral orders, while the signal of SiO is localised to the fundamental band at 4$\mu$m (with a minimal contribution from order 2 at 4.75 $\mu$m), and the CO originates predominately from the noisy 4.75 $\mu$m order, with minor contribution from the 5.2 $\mu$m order which suffers from extreme thermal background. The combined model, denoted `All' is detected at a S/N of 21.6.

\subsection{Atmospheric retrievals}

\begin{figure*}
    \includegraphics[width=2\columnwidth]{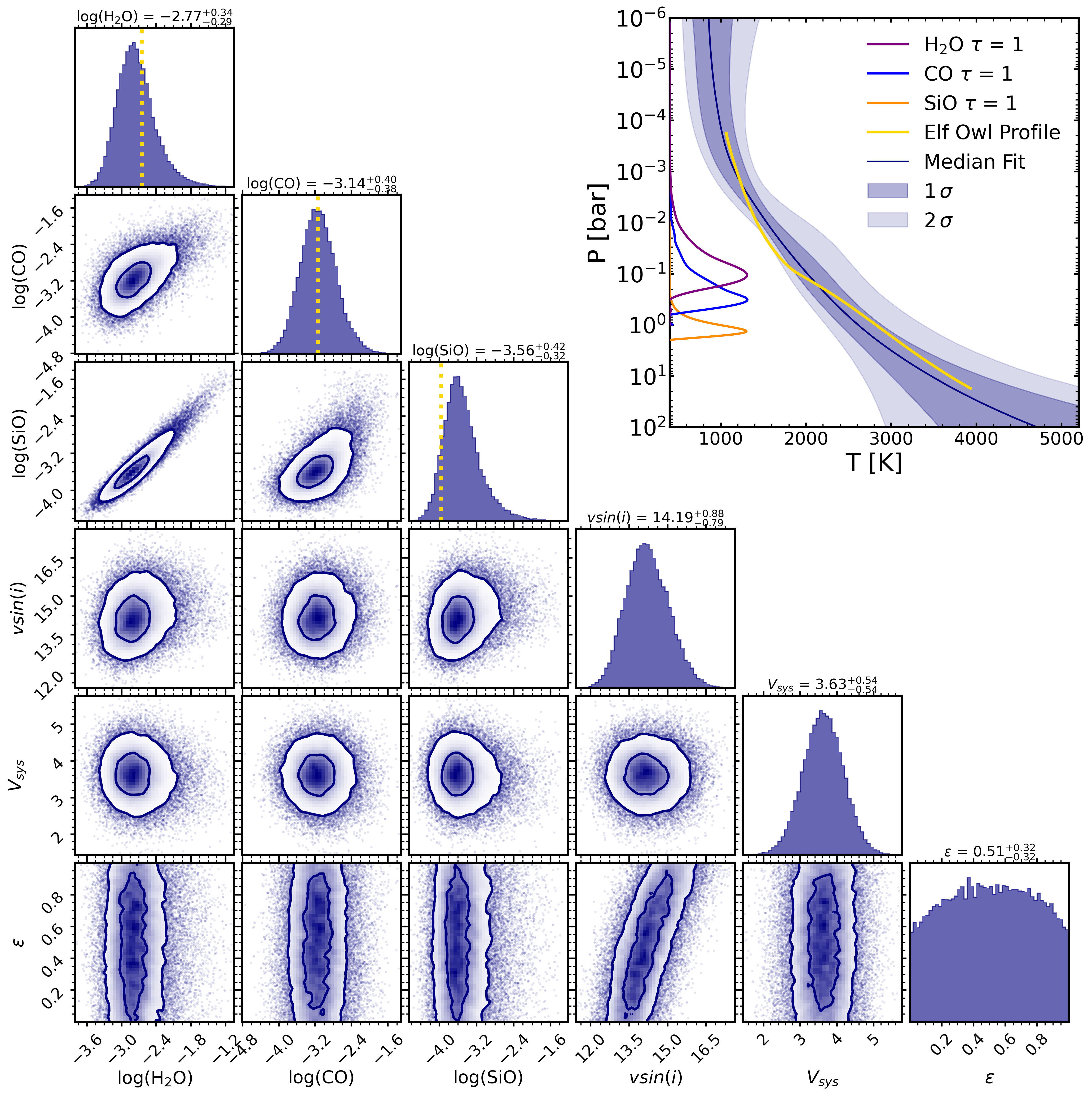}
    \caption{Retrieved chemical abundances and dynamics parameters from our retrieval. Note the degeneracy between the absolute abundance measurements of each molecular species, most notably between SiO and H$_2$O, the main source of uncertainty when inferring absolute abundances and metallicity measurements. The predicted abundances of H$_2$O, CO, and SiO from the Sonora Elf Owl model that best fits the properties of TWA\,5\,B (T$_{\text{eff}}$= 2400 K, log(g) = 4, [M/H]$_{\odot}$ = 0.5, [C/O]$_{\odot}$= 0.5, log(K$_{zz}$) = 8.0) are plotted in gold. Top right: the retrieved TP profile, following the adopted five-knot parametrisation. $\tau$ = 1 surfaces for each of the retrieved molecules are plotted, demonstrating the maximum depth probed by these observations, with SiO probing deeper in the atmosphere than CO and H$_2$O. The profile from the best fitting Sonora Elf Owl model is plotted in gold.}
    \label{fig:big_corner_with_TP}
\end{figure*}

To measure the abundances of the detected molecules, we carry out a Bayesian retrieval using CCF to Log-L mapping, adopting the likelihood framework of \citet{Brogi2019}. We use the HyDRA framework (see Section \ref{sec:models_and_retreivals}), and perform a free chemistry retrieval to measure the abundances of each of the three detected species, alongside the atmospheric dynamics and temperature-pressure profile (Figure \ref{fig:big_corner_with_TP}; see Table \ref{tab:combined_table}). We retrieve H$_2$O to be the highest abundance volatile molecule in the atmosphere of TWA\,5\,B, with a volume mixing ratio log(H$_2$O) = -2.77$^{+0.34}_{-0.29}$. The gas-phase SiO abundance is well constrained to less than half a dex precision, with a measured abundance of log(SiO) =-3.56$^{+0.42}_{-0.32}$, while CO is retrieved with an abundance of log(CO) = $-3.14^{+0.40}_{-0.38}$. The abundances of SiO and H$_2$O show a strong degeneracy due to the relative impact of the two opacities at 4 $\mu$m. As the SiO lines form deeper in the atmosphere than H$_2$O (Figure \ref{fig:big_corner_with_TP}), increasing the H$_2$O abundance will obscure the SiO lines in the model, requiring a higher SiO abundance to match the strength of the SiO in the data. As the spectral lines of each molecule form at different pressure levels, this correlation occurs between each of the retrieved abundances, and is the main source of uncertainty on the retrieved molecular abundances.

\begin{table}
	\centering
	\caption{Retrieved posterior distributions and corresponding priors for all model parameters. \textit{U} denotes uniform priors, while \textit{N}($\mu$, $\sigma$) denotes Gaussian priors with a normal distribution with mean $\mu$ and standard deviation $\sigma$. \label{tab:combined_table}}
	\renewcommand{\arraystretch}{1.3}
	\begin{tabular}{lcc}
		\hline
		Parameter & Posterior & Prior range \\
		\hline
		\hline
		
		\multicolumn{3}{l}{\textbf{Chemistry}} \\
		log(H$_2$O [VMR]) & $-2.77^{+0.34}_{-0.29}$ & \textit{U}(-12 $\rightarrow$ -1) \\
		log(CO [VMR]) & $-3.14^{+0.40}_{-0.38}$ & \textit{U}(-12 $\rightarrow$ -1) \\
		log(SiO [VMR]) & $-3.56^{+0.42}_{-0.32}$ & \textit{U}(-12 $\rightarrow$ -1) \\
		
		\\[-0.5em]
		\multicolumn{3}{l}{\textbf{Dynamics}} \\
		v$_{\text{sys}}$ [km/s] & $3.64^{+0.49}_{-0.48}$ & \textit{U}(-10 $\rightarrow$ 10) \\
		$v\,\sin(i)$ [km/s] & $14.19^{+0.88}_{-0.79}$ & \textit{U}(1 $\rightarrow$ 50) \\
		$\epsilon$ & $0.51^{+0.32}_{-0.32}$ & \textit{U}(0 $\rightarrow$ 1) \\
		
		\\[-0.5em]
		\multicolumn{3}{l}{\textbf{Companion parameters}} \\
		R$_p$ [R$_J$] & $2.65^{+0.68}_{-0.71}$ & \textit{N}(2.7, 0.8) \\
		$\log$(g [cm/s$^2$]) & $4.01^{+0.18}_{-0.18}$ & \textit{N}(3.9, 0.2) \\
		
		\\[-0.5em]
		\multicolumn{3}{l}{\textbf{TP Profile}} \\
		T$_{set}$ [K] & $1514^{+733}_{-294}$ & \textit{U}(1000 $\rightarrow$ 4000) \\
		log(P$_{set}$ [bar]) & $-2.09^{+1.6}_{-0.57}$ & \textit{U}(-3 $\rightarrow$ 1) \\
		log(dP$_{set}$ top [bar]) & $2.03^{+0.64}_{-0.83}$ & \textit{U}(0.5 $\rightarrow$ 3) \\
		log(dP$_{set}$ bottom [bar])  & $0.74^{+0.16}_{-0.16}$ & \textit{U}(0.5 $\rightarrow$ 1) \\
		$\alpha_{0}$ & $-0.134^{+0.087}_{-0.096}$ & \textit{U}(-0.3 $\rightarrow$ 0) \\
		$\alpha_{1}$ & $-0.103^{+0.042}_{-0.043}$ & \textit{U}(-0.3 $\rightarrow$ 0) \\
		$\alpha_{2}$ & $-0.180^{+0.060}_{-0.072}$ & \textit{U}(-0.3 $\rightarrow$ 0) \\
		$\alpha_{3}$ & $-0.045^{+0.080}_{-0.043}$ & \textit{U}(-0.1 $\rightarrow$ 0.1) \\
		$\alpha_{4}$ & $-0.008^{+0.068}_{-0.061}$ & \textit{U}(-0.1 $\rightarrow$ 0.1) \\
		
		\\[-0.5em]
		\multicolumn{3}{l}{\textbf{Clouds}} \\
		log($\kappa$ [cm$^2$/g]) & $-7.0^{+9.6}_{-8.5}$ & \textit{U}(-20 $\rightarrow$ 50) \\
		log($\gamma_{\text{cl}}$) & $11.0^{+6.0}_{-6.8}$ & \textit{U}(0 $\rightarrow$ 20) \\
		log(P$_{\text{cl}}$ [bar]) & $-1.3^{+2.4}_{-3.1}$ & \textit{U}(-6 $\rightarrow$ 2) \\
		
		\hline
		\hline
	\end{tabular}
\end{table}

The abundances of the chemical species in emission spectroscopy are inherently coupled to the adopted temperature pressure parametrisation. We retrieve a well constrained TP profile ($\pm$ 200 K) at the pressures we are directly probing in our observations (-2 $\lesssim$ log(P [bar]) $\lesssim$ 0), while the upper and deep atmosphere remain unconstrained by the data, and restricted only by the choice of priors. The data is therefore not sensitive to the presence of temperature inversions in the upper atmosphere, as has been observed in some isolated brown dwarfs (e.g. \citealt{Nasedkin2025}). An inverted upper atmosphere remains a possibility for TWA\,5\,B, and would be consistent with the upper atmosphere heating or accretion shocks required to produce the soft X-ray spectrum observed in \citet{Tsuboi2003}. The observations are maximally sensitive to log(P [bar]) = -1 to log(P [bar]) = -2 bar pressures through the strong and panchromatic H$_2$O opacity, with the 4 $\mu$m SiO lines probing deeper, down to photospheric pressures of log(P [bar]) = 0. We note that the uncertainties on the full TP profile at any given pressure (Figure \ref{fig:big_corner_with_TP}) are lower than the uncertainty in the posterior distributions of individual TP profile parameters in the retrieval (e.g. T$_\mathrm{set}$) due to the correlations between the TP profile parameters (see Figure \ref{fig:Big_corner_plot}), which tightly constrain the global profile. Our retrieved temperature profile shows a good match to theoretical models of M8.5 dwarfs, and in Figure \ref{fig:big_corner_with_TP} we plot the retrieved TP profile against a Sonora Elf Owl model \citep{Mukherjee2024_elfowl} matching the observed properties and chemistry of TWA\,5\,B (T$_{\text{eff}}$= 2400 K, log(g) = 4, [M/H]$_{\odot}$ = 0.5, [C/O]$_{\odot}$= 0.5, log(K$_{zz}$) = 8.0). Our retrieved TP profile is consistent within 2$\sigma$. The retrieved radius (R$_{\text{p}}$) of the companion, and its surface gravity $\log(g)$ have prior dominated posterior distributions (see Section \ref{sec:models_and_retreivals}). The full corner plot from the retrieval can be found in Figure \ref{fig:Big_corner_plot}.

\subsubsection{Dynamics}

The spectral lines of the companion are broadened by the projected rotational velocity of TWA\,5\,B, which we measure to have a vsin(i) of 14.19$^{+0.88}_{-0.79}$ km/s. This measurement is a refinement of the previously reported 16 $\pm$ 2 km/s for TWA\,5\,B \citep{Mohanty2003}. The projected rotational velocity suffers from a weak degeneracy with the companion's limb darkening, parametrised by the coefficient $\epsilon$ (see Figure \ref{fig:big_corner_with_TP}). The limb darkening is poorly constrained, but shows a very weak preference for values ($\sim$ 0.5), implying a relatively smooth centre-to-limb intensity gradient. The measured orbital velocity (v$_{\text{sys}}$) is 3.63$\pm0.54$ km/s, which deviates from predictions of the planetary orbital radial velocity, calculated to be 0.3 $\pm$ 0.7 km/s using the ephemeris provided in \citet{Bowler2020}, using the whereistheplanet\footnote{\url{http://whereistheplanet.com/}} prediction tool \citep{Wang2021b}. While this offset could be driven by a discrepancy in the orbital eccentricity of TWA\,5\,B, the offset is consistent with the predicted value when considering the uncertainties on the stellar systemic velocity, 10.2 $\pm$ 2.2 km/s \citep{Riedel2017}. It is therefore challenging to constrain the radial motion of the orbit of TWA\,5\,B from this single data point. The binary motion of the M-dwarf hosts complicates further refinement of the stellar systemic velocity using our data.

\subsubsection{Isotopologues}
\label{sec:Isotopes}

The fundamental band of CO in the \textit{M}-band has been identified to provide the strongest features of $^{13}$CO absorption \citep{Molliere2019b}. We therefore additionally search for the carbon monoxide isotopologue $^{13}$CO in our data, but see limited evidence (S/N $<$ 3) from the cross-correlation function. We therefore do not include the atmospheric $^{13}$CO abundance as a free parameter in our main free chemistry retrieval (Section \ref{sec:models_and_retreivals}). To assess the impact of this modelling choice we run an additional test of running a separate retrieval with the $^{13}$CO as an additional parameter, independent of the $^{12}$CO abundance used in the main retrieval. In this case, despite the non-detection in cross-correlation, we obtain a well constrained posterior distribution with log($^{13}$CO) = $-3.68^{+0.61}_{-0.56}$, alongside the measurement of log($^{12}$CO) = $-3.08^{+0.40}_{-0.39}$ (consistent with the measurement of CO in the main retrieval). However, using the relation 

\[\mathrm{^{12}C/^{13}C} = \frac{\mathrm{X_{^{12}CO}}}{\mathrm{X_{^{13}CO}}},\] 

this leads to an exceptionally low and unphysical isotope fraction $^{12}$C/$^{13}$C $\sim$ 4, compared to the solar value (93.5 $\pm$ 3; \citealt{Lyons2018}), and the local ISM (68 $\pm$ 5; \citealt{Milam2005}). In cases where the evidence from the cross-correlation is not decisive, retrieval posteriors for $^{12}$C/$^{13}$C appear to be dominated by modelling assumptions rather than reflecting a strong physical detection. There is evidence to suggest that in lower signal to noise spectra the $^{12}$C/$^{13}$C ratio tends towards very low values, as retrieval frameworks will converge on high abundances of $^{13}$CO due to aliasing between the similar $^{13}$CO and $^{12}$CO spectral features. This has been seen in the directly imaged planet YSES 1 b, in which lower signal to noise studies that supported a low $^{12}$C/$^{13}$C ratio of 31$^{+17}_{-10}$ \citep{Zhang2021a}, were revised to the higher and more physically motivated value of 88$\pm$13 with higher signal to noise data \citep{Zhang2024}. We propose that the same effect impacts our results, which are retrieving the $^{13}$CO abundance predominantly from the thermal background dominated and low signal to noise 4.7 $\mu$m order, and that the inference of $^{13}$CO from the retrieval does not reflect an accurate abundance measurement of $^{13}$CO. Further observations of TWA\,5\,B as part of the SupJup survey \citep{deRegt2024} will provide improved sensitivity to $^{13}$CO through high signal to noise observations in the \textit{K}-band.
 
\section{Discussion}
\label{sec:Discussion}

\subsection{The detection of SiO in a directly imaged companion}

We detect a strong signature of gaseous SiO, at S/N = 7.5, marking the first such conclusive detection of gaseous SiO in a directly imaged companion, and confirming predictions of silicon monoxide (SiO) as the most abundant silicon-bearing gas in the atmospheres of hot young giant planets \citep{Sharp2007, Visscher2010}. Previous observations of gaseous SiO in sub-stellar atmospheres have proved challenging, as the strongest SiO ro-vibrational band heads of SiO fall in the near- to mid-infrared at 4 $\mu$m (first-overtone, $\Delta$v = 2) and 9 $\mu$m (fundamental band), both of which are challenging regimes to obtain the required high signal to noise and high-resolution spectra. In directly imaged systems a tentative signal of SiO has been observed in $\beta$ Pic b using CRIRES+ \textit{M}-band spectroscopy \citep{Parker2024}, and SiO has been detected in isolated brown dwarfs with JWST \citep{GonzalezPicos2025b}, while \citet{Molliere2025} detect SiO grains at 10 $\mu$m in the isolated sub-stellar object PSO J318.5338-22.8603. For transiting systems a detection of the 4 $\mu$m SiO feature in the ultra-hot Jupiter WASP-121~b has been presented using JWST/G395H spectra \citep{Evans-Soma2025, Gapp2025}, while SiO has additionally been proposed to explain the UV and NIR spectrum of WASP-178 b \citep{Lothringer2022, Lothringer2025}.

\begin{figure}
    \includegraphics[width=\columnwidth]{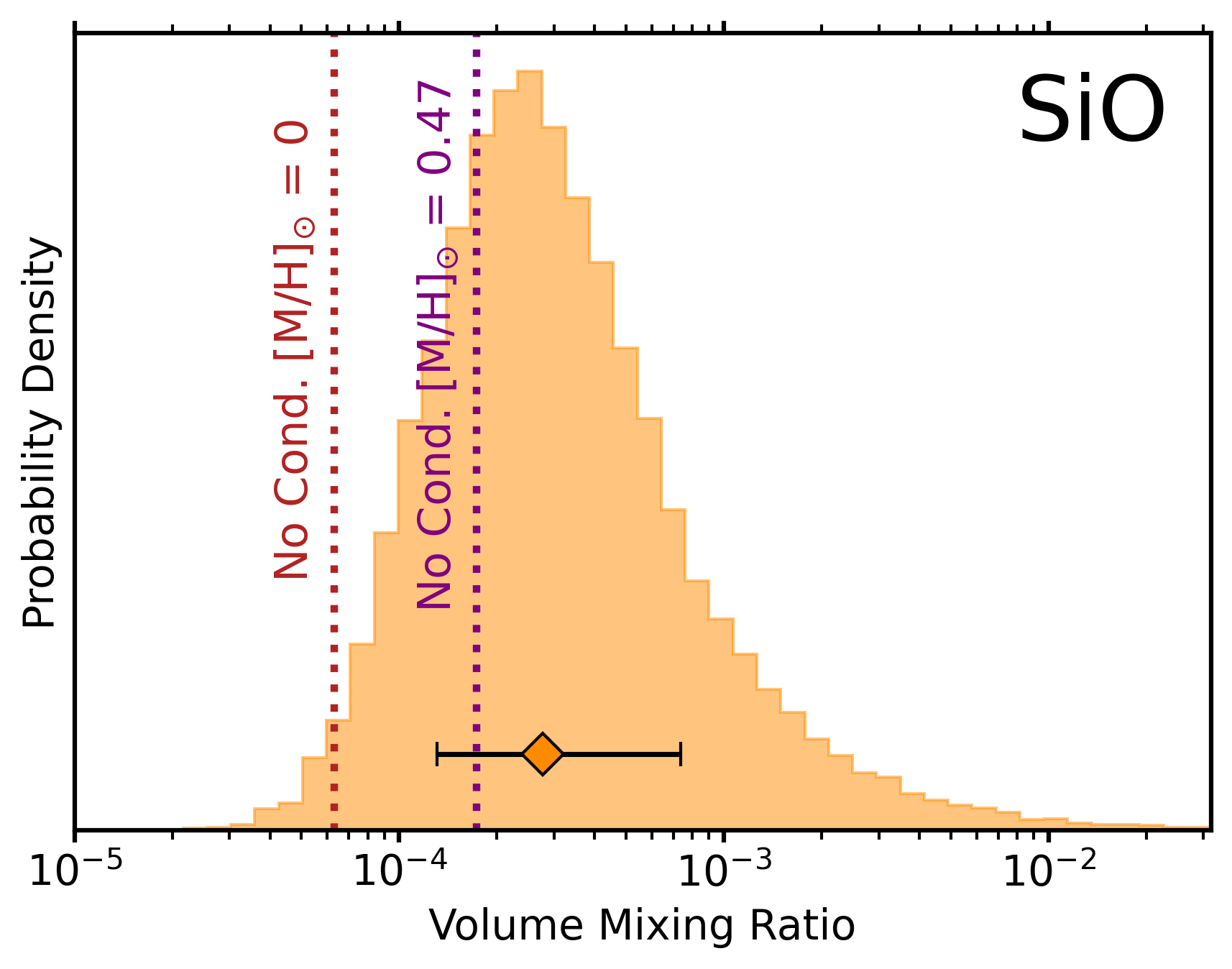}
    \caption{Retrieved posterior for the absolute abundance of SiO in volume mixing ratio (VMR) in TWA\,5\,B, log(SiO) = -3.56$^{+0.42}_{-0.32}$. The dotted purple line denotes the predicted SiO abundance at the retrieved metallicity ([M/H]$_{\odot}$ = $0.47^{+0.34}_{-0.29}$), in the case of no magnesium-silicate condensation, matching the retrieved atmospheric abundance of SiO, and implying the absence of magnesium silicate clouds at or below the atmospheric pressures of TWA\,5\,B probed in this work. The red line denotes the predicted abundance of SiO in a solar metallicity atmosphere in the absence of magnesium-silicate condensates. The largest uncertainty in this absolute abundance measurement arises from degeneracies with other molecular abundances, primarily the relative impact of SiO and H$_2$O opacity at 4 $\mu$m.}
    \label{fig:SiO_abund}
\end{figure}

\subsection{SiO abundance as a diagnostic of cloud condensation}

The gas chemistry and abundance of SiO in hot giant planets is influenced primarily by silicate and magnesium-silicate cloud condensates, and SiO will most readily condense to form Mg$_2$SiO$_4$ (forsterite), MgSiO$_3$ (enstatite) or SiO$_2$ (quartz), providing an atmospheric sink of gaseous SiO \citep{Gao2021}. The theoretical predictions on the evolution of the atmospheric silicon abundance from gas phase to condensates are supported by the measurement of the silicate absorption index across populations of isolated brown dwarfs with the Spitzer Infrared Spectrograph \citep{Suarez2022}, which imply a shift from clear to cloudy atmospheres at effective temperatures of $\sim$2000 K, roughly corresponding to the M/L spectral type transition. The abundance of gaseous SiO above a magnesium-silicate cloud deck follows the saturation pressure abundance, while below a cloud deck, or in the absence of magnesium-silicate clouds, SiO is predicted to be the most abundant silicon-bearing gas in the atmosphere, with an abundance of log(SiO) $\approx$ -4.20 + [M/H] \citep{Visscher2010}.

Here, two lines of evidence from our retrieved parameters imply that the atmosphere of TWA\,5\,B globally lacks magnesium silicate condensation. Firstly, we retrieve a strong signal of gaseous SiO with an abundance of log(SiO)~=~-3.56$^{+0.42}_{-0.32}$ VMR (Figure \ref{fig:SiO_abund}) at pressures of -2 $\lesssim$ log(P [bar]) $\lesssim$ 0. This excludes the presence of magnesium silicate clouds at these pressures as the formation of condensates would significantly deplete the abundance of gas phase SiO as silicon is sequestered from the gas phase into condensates \citep{Visscher2010}. Furthermore, this additionally implies that all atmospheric layers deeper than the pressures probed (i.e. log(P [bar])~$\gtrsim$~0) must also be free of magnesium-silicate clouds, as the formation of deep cloud layers would deplete the overlaying atmosphere of gas-phase silicon through rainout condensation. 

Second, the constraints on the cloud parameters obtained in the retrieval provide further evidence to support the absence of magnesium-silicate cloud condensation. We retrieve a very low cloud opacity (log($\kappa$ [cm$^2$ g$^{-1}$]) = $-7.0^{+9.6}_{-8.5}$), indicating that cloud opacities have a minimal impact on the observed spectrum. Furthermore, the retrieved cloud deck pressure is poorly constrained and pushed towards higher pressures (log(P$_{\text{cl}}$ [bar]) = $-1.3^{+2.4}_{-3.1}$), disfavouring significant cloud absorption. These retrieved parameters indicate that there is no discernible spectroscopic impact of cloud opacity on the observed M-band spectrum of TWA\,5\,B, disfavouring the presence of high-altitude clouds.

The global absence of magnesium-silicate cloud condensation on TWA\,5\,B is therefore inferred from the abundance measurement of SiO, which provides a direct probe of the condensation chemistry by accessing the cloud-precursor species silicon in the gas phase, and supported by the retrieved absence of cloud opacity impacting the observed spectrum. This conclusion supports predictions for condensation in sub-stellar atmospheres, as the $\sim$2400 K equilibrium temperature of TWA\,5\,B \citep{Neuhauser2000, Bonnefoy2014} lies above the proposed temperature transition from absent to thick magnesium-silicate clouds in the isolated brown dwarf population. 

An alternative interpretation that is consistent with the observed SiO abundance is that the atmosphere of TWA\,5\,B possesses high-altitude magnesium-silicate clouds which are transparent at 4 $\mu$m, permitting our observations to probe the equilibrium abundances of SiO below the cloud deck at $\sim$~1~bar pressures. Following Mie scattering in the small-particle limit, for a cloud deck to be transparent at 4 $\mu$m would require a cloud deck of <~0.1~$\mu$m grains \citep[e.g.][]{Taylor2021}. However, high-altitude clouds are disfavoured by the temperature and spectral class of TWA\,5\,B, the absence of cloud features at shorter wavelengths \citep{Neuhauser2000, Bonnefoy2014}, and the lack of evidence for cloud opacity impacting our spectra.

\subsection{SiO as a probe of the refractory content of giant planet atmospheres}

\begin{figure}
    \includegraphics[width=\columnwidth]{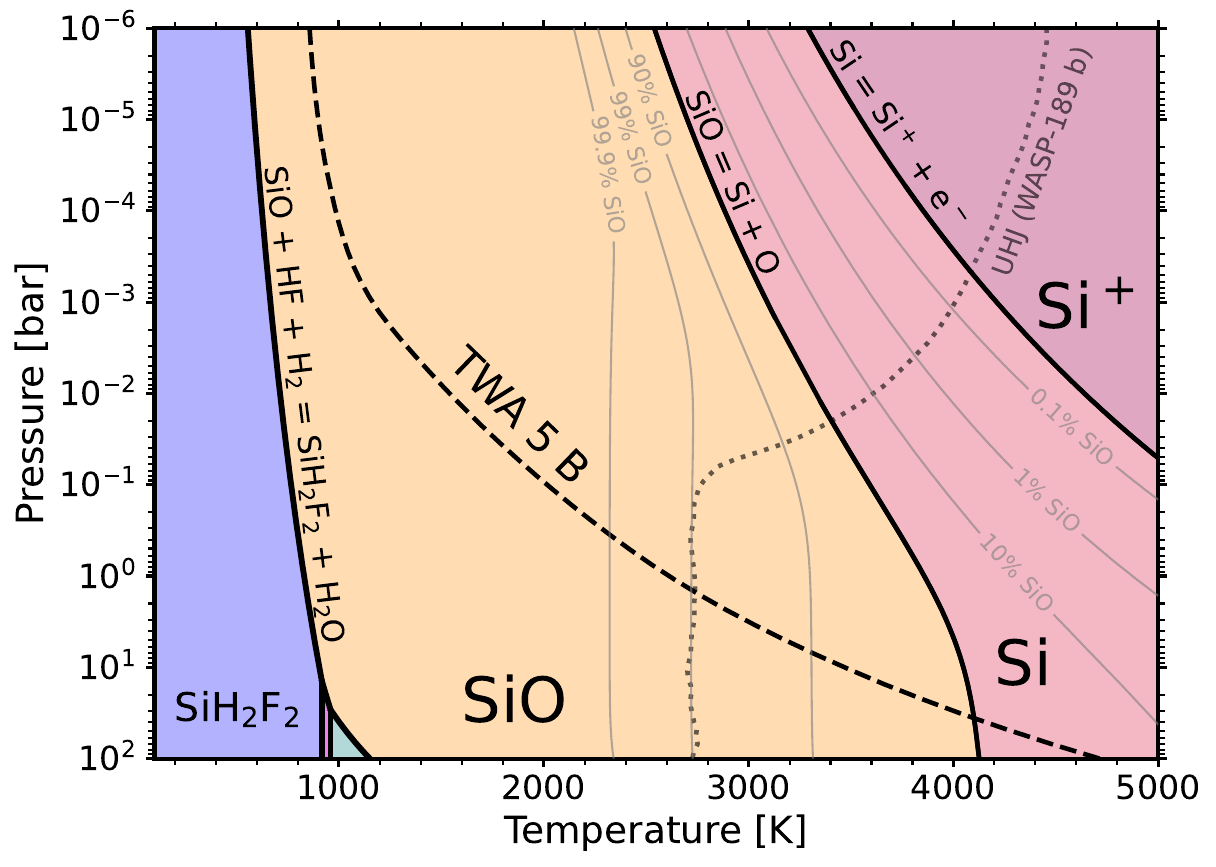}
    \caption{Overview of the dominant silicon-bearing species in giant planet atmospheres as a function of temperature and pressure, with no condensation. Solid lines indicate where major silicon-bearing gases have equal abundances, including the chemical transition from SiH$_2$F$_2$ to SiO at lower temperatures, the high temperature dissociation of SiO into atomic silicon (Si, also denoted as Si\,\textsc{i}), and the ionisation of atomic silicon to Si$^{+}$ (Si\,\textsc{ii}). The pink and green regions of the plot indicate SiH$_3$F and SiH$_4$, respectively, important high pressure carriers of silicon \citep{Faherty2025}. Grey contours indicate the fraction of the atmospheric silicon contained within molecular SiO across the dissociation boundary. The retrieved TP profile for TWA\,5\,B is plotted (dashed line), demonstrating that SiO is the dominant silicon-bearing species in TWA\,5\,B. The TP profile for the ultra-hot Jupiter WASP-189 b (dotted) is shown for reference \citep{Sanchez2025}.
    }
    \label{fig:Si_dissociation}
\end{figure}

A further consequence of the implied absence of magnesium-silicate cloud species in the atmosphere of TWA\,5\,B is that the entire atmospheric abundance of silicon likely remains locked in the gas phase, and almost exclusivity in SiO, with minimal silicon entrained in gas-phase SiS ($\sim$ 1 per cent). The strength of the Si-O bond ensures that the SiO abundance is not significantly depleted by dissociation in the observed atmosphere of TWA\,5\,B, with $<$ 1 per cent of the total silicon atmospheric silicon held in neutral atomic silicon (Si\,\textsc{i}) at the 1 bar pressures probed by the detection of SiO in TWA\,5\,B. This is unlike the case of ultra hot Jupiters in which atomic and ionised silicon dominate the observed silicon budget \citep[e.g.][]{Sanchez2025}, produced through the dissociation and ionisation of SiO in the strongly inverted upper atmosphere. In Figure \ref{fig:Si_dissociation}, we plot the dominant silicon-bearing species in giant planet atmospheres as a function of pressure and temperature, with the solid lines indicating where major silicon-bearing gases have equal abundances. We assume a solar metallicity gas and C/O, and calculate the chemistry with \textsc{FastChemCond} \citep{Kitzmann2024}, following \citep{Visscher2010}. The observed TP profile for TWA\,5\,B is over plotted, demonstrating that SiO is the dominant silicon-bearing species in the observable atmosphere of TWA\,5\,B above 10 bar.

Therefore, through the abundance measurement of SiO we are directly probing the total abundance of refractory silicon in the envelope of TWA\,5\,B at $\sim$1 bar pressures. Under the assumption of a well mixed atmosphere, this SiO abundance is a pristine measurement of the initial Si abundance of TWA\,5\,B at formation and therefore, the atmospheric ratios of refractory (Si) to volatile (C, O) elements in TWA\,5\,B can trace the processes through which the sub-stellar companion formed \citep{Chachan2023, Lothringer2021}.

\begin{figure*}
    \includegraphics[width=2\columnwidth]{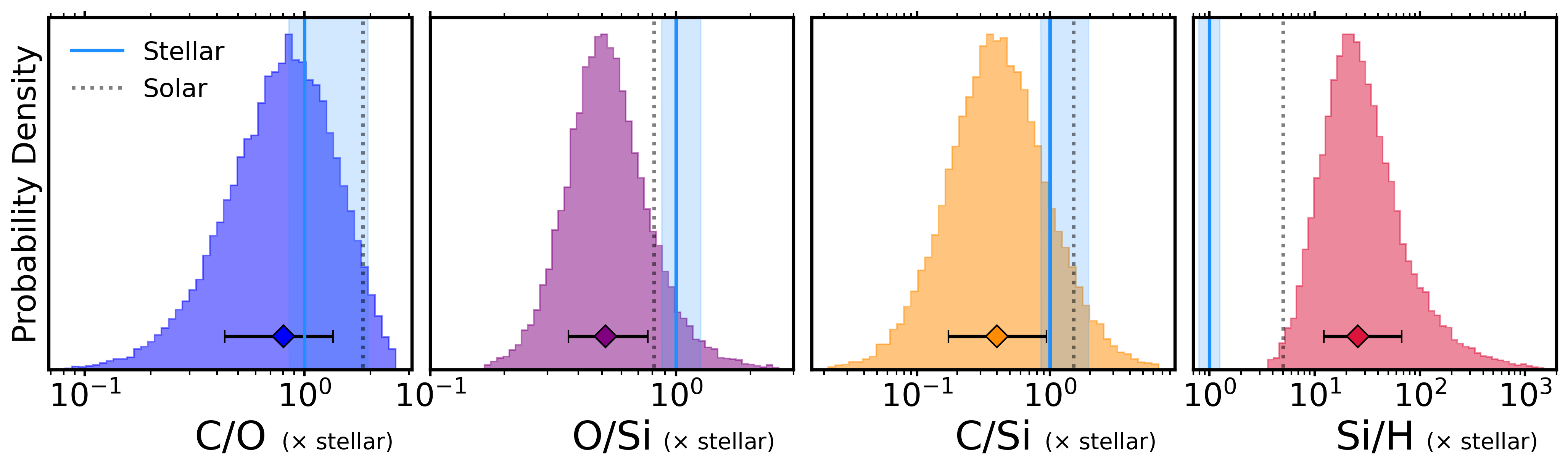}
    \caption{Measured C/O and volatile-to-refractory abundance ratios for TWA\,5\,B presented relative to the stellar values (blue, with shaded 1$\sigma$ uncertainties). These results are compared to the solar abundances (grey dotted; calculated from \citealt{Asplund2021}). We measure a stellar C/O ([C/O]$_{\star}$ =-0.09$^{+0.23}_{-0.27}$), and sub-stellar O/Si ([O/Si]$_{\star}$ = $-0.29^{+0.17}_{-0.15}$) and C/Si ([C/Si]$_{\star}$ = $-0.40^{+0.37}_{-0.36}$), but a super-stellar Si/H ([Si/H]$_{\star}$ = $1.41^{+0.42}_{-0.32}$), indicating a significant enrichment of silicon in the atmosphere of TWA\,5\,B relative to both the solar value and the stellar abundance. The sub-stellar O/Si and C/Si is driven primarily by the silicon enrichment of TWA\,5\,B, rather than a depletion of volatile elements. 
    }
    \label{fig:refract_ratios}
\end{figure*}

Here, we report volatile-to-refractory ratios for a directly imaged companion using the measured abundance of SiO. We calculate the ratio Si/H, which traces the ratio of gaseous to solid components accreted during formation, and the O/Si and C/Si ratios which trace the ratio of rocks to ice components \citep{Lothringer2021, Chachan2023}. The abundance ratios are defined as follows:

\[
\mathrm{C/O} = \frac{\mathrm{X_{CO}}}{\mathrm{X_{CO}} + \mathrm{X_{H_{2}O}} + \mathrm{X_{SiO}}}, \, \, \, \, \, \, \, \, \, \, \, \, \mathrm{Si/H} = \frac{\mathrm{X_{SiO}}}{2\,\mathrm{X_{H_{2}}} + 2\,\mathrm{X_{H_{2}O}}},\]
\[ \mathrm{O/Si} = \frac{\mathrm{X_{CO}} + \mathrm{X_{H_{2}O}} + \mathrm{X_{SiO}}}{\mathrm{X_{SiO}}}, \, \, \, \, \, \, \, \, \, \, \, \, \mathrm{C/Si} = \frac{\mathrm{X_{CO}}}{\mathrm{X_{SiO}}}.
\]

The observed elemental ratios of C/O (0.26$^{+0.17}_{-0.12}$), O/Si (9.6$^{+4.7}_{-2.8}$), and C/Si (2.4$^{+3.2}_{-1.3}$) are all depleted compared to the Sun, with the exception of Si/H ($1.66^{+2.69}_{-0.87}\times10^{-4}$) which is significantly enriched relative to solar abundances (Table \ref{tab:refractories_table}). The solar abundance ratios are calculated from the solar photospheric abundances of oxygen, carbon, and silicon presented in \citet{Asplund2021}, and have precisions of $\leq$ $\pm$0.04 dex. 

However, the abundance ratios in the atmosphere of TWA\,5\,B must be interpreted within the context of the host star. In the following discussion, it is important to caveat that the only existing measurements of stellar abundances for TWA\,5 are derived from X-ray spectra from XMM-Newton \citep{Argiroffi2005}. These provide precise abundances, but are crucially measurements of the coronal plasma, as opposed to the photospheric pressures probed in optical spectroscopy. These abundances can therefore be modified due to the higher temperatures and active accretion, which drives the X-ray flux. Very few optical spectroscopy stellar abundance measurements exist for TW Hya moving group stars, but broadly show sub-solar to solar metallicities \citep{Laskar2009, Donati2024}. In the discussion that follows, we will principally report abundance ratios relative to the measured stellar composition of TWA\,5 A, with the logarithmic abundance ratio relative to stellar abundances denoted as [X/Y]$_{\star}$. We will also report abundances relative to solar for reference, denoted as [X/Y]$_{\odot}$ for clarity.

\begin{table}
	\centering
	\caption{Details of the calculated chemical abundance ratios, inferred from the retrieval. Square brackets denote the logarithmic abundance ratios relative to stellar ([X/Y]$_{\star}$) and solar ([X/Y]$_{\odot}$) abundances. \label{tab:refractories_table}}
    \renewcommand{\arraystretch}{1.3}
	\begin{tabular}{cccc} 
		\hline
       Parameter & Absolute & [X/Y]$_{\star}$ & [X/Y]$_{\odot}$ \\
		\hline
            \hline
            M/H &
            $2.3^{+2.9}_{-1.2}\times10^{-3}$ &
            $1.15^{+0.34}_{-0.29}$ &
            $0.47^{+0.34}_{-0.29}$ \\
            
            O/H &
            $1.72^{+2.01}_{-0.81}\times10^{-3}$ &
            $1.15^{+0.34}_{-0.28}$ &
            $0.55^{+0.34}_{-0.28}$ \\
            
            C/H &
            $4.4^{+6.6}_{-2.6}\times10^{-4}$ &
            $1.06^{+0.40}_{-0.38}$ &
            $0.18^{+0.40}_{-0.38}$ \\
            
            C/O &
            $0.26^{+0.17}_{-0.12}$ &
            $-0.09^{+0.23}_{-0.27}$ &
            $-0.36^{+0.23}_{-0.27}$ \\
            
            O/Si &
            $9.6^{+4.7}_{-2.8}$ &
            $-0.29^{+0.17}_{-0.15}$ &
            $-0.20^{+0.17}_{-0.15}$ \\
            
            C/Si &
            $2.4^{+3.2}_{-1.3}$ &
            $-0.40^{+0.37}_{-0.36}$ &
            $-0.57^{+0.37}_{-0.36}$ \\
            
            Si/H &
            $1.66^{+2.69}_{-0.87}\times10^{-4}$ &
            $1.41^{+0.42}_{-0.32}$ &
            $0.71^{+0.42}_{-0.32}$ \\

		\hline
  \hline
	\end{tabular}
\end{table}

XMM-Newton has measured TWA\,5 A to be metal poor, with varying levels of depletion of individual elements \citep{Argiroffi2005}. Using measured stellar abundances of silicon, oxygen, and carbon, we determine the stellar O/Si, and C/Si ratios to be approximately solar, but that the host star has a sub-solar C/O, and is depleted in silicon relative to the Sun. For TWA\,5\,B we therefore find the sub-solar C/O measurement (C/O = $0.26^{+0.17}_{-0.12}$) to be consistent with the absolute stellar C/O (C/O$_\mathrm{TWA\,5\,A}$ $\sim$ 0.32), with [C/O]$_{\star}$ =-0.09$^{+0.23}_{-0.27}$. The atmospheric O/Si and C/Si measurements for TWA\,5\,B are slightly sub-stellar, but consistent within $2\sigma$ of the stellar values (Figure \ref{fig:refract_ratios}). The sub-stellar O/Si and C/Si driven primarily by the silicon enrichment of TWA\,5\,B, rather than a depletion of volatile elements, and we observe super-solar C/H and O/H. The largest discrepancy from the stellar value is the super-stellar Si/H ratio, with [Si/H]$_{\star}$ = $1.41^{+0.42}_{-0.32}$ (Figure \ref{fig:refract_ratios}). Assuming a well mixed atmosphere, this represents a $\sim$25 times enrichment of silicon in TWA\,5\,B relative to the stellar value, and is discrepant from the stellar value by 4$\sigma$, and the solar silicon abundance by 2$\sigma$. 

However, the Si/H ratio is dependent on the measurement of a single retrieved molecular abundance, the atmospheric SiO VMR. We must therefore consider whether this enrichment of silicon in TWA\,5\,B relative to the stellar value could arise from chemical disequilibrium driving an enhanced SiO abundance. At the pressures (-2 $\lesssim$ log(P [bar]) $\lesssim$ 0) and temperatures (1800 K $\lesssim$ T $\lesssim$ 2700 K) probed in TWA\,5\,B, SiO is predicted to be the dominant silicon-bearing molecule in equilibrium (Figure \ref{fig:Si_dissociation}), containing $>99$ per cent of the atmospheric silicon abundance \citep{Visscher2010}. Given that SiO already holds almost the entire reservoir of atmospheric silicon in equilibrium, disequilibrium gas phase chemistry is unlikely to be able to lead to an enhanced SiO abundance. Furthermore, the atmospheric layer probed in this work represents the maximum predicted equilibrium abundance of SiO in TWA\,5\,B, with higher pressure and temperature regions predicted to have lower abundances of SiO due to thermal dissociation and chemical conversion to SiH$_4$, disfavouring deep vertical mixing as a route to enhance the observed SiO abundance. Therefore, to produce the observed SiO abundance in TWA\,5\,B requires an enrichment of the atmosphere with silicon (i.e. a super-stellar Si/H). At the temperatures and pressures probed in this work, SiO appears to be an excellent probe of the solid enrichment of TWA\,5\,B, as its maximum abundance is primarily set by the bulk silicon enrichment of the atmosphere.

\begin{figure*}
    \includegraphics[width=2\columnwidth]{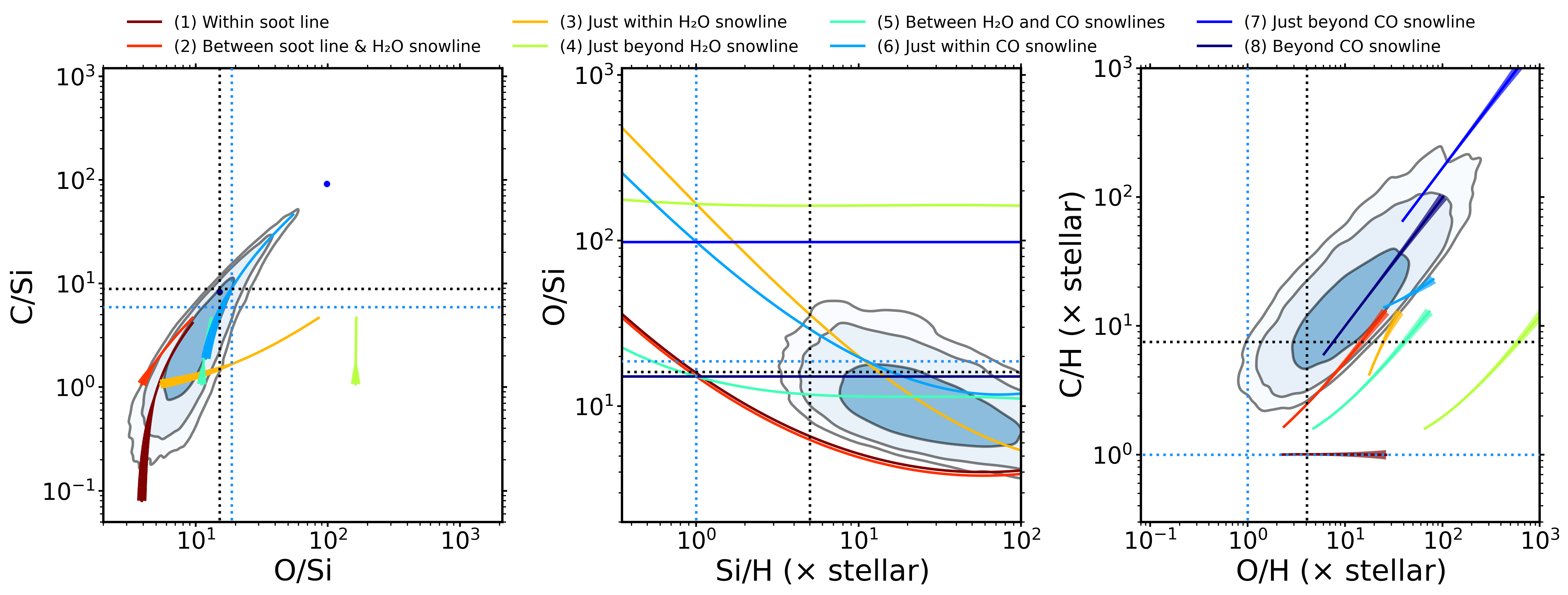}
    \caption{Retrieved posterior contours compared with the formation models from \citep{Chachan2023}, tracing the expected elemental abundance ratios from core accretion scenarios in different regions of a protoplanetary disc. The dotted blue lines denote the stellar abundance ratios, which is the first order expected composition of TWA\,5\,B in the case of a star-like gravitational instability formation pathway. The solar abundance ratios are shown in black dotted lines. We show the three sigma confidence intervals, and the models from \citet{Chachan2023} in the left and right panels have been truncated to contain only values that are consistent within 3$\sigma$ with the observed Si/H ratio, with the width of the model lines corresponding to the Si/H ratio (thick: high Si/H, thin: low Si/H). These diagnostic plots demonstrate that formation just within the CO snowline and formation beyond the CO snowline are both consistent with the observations within 2$\sigma$. Following \citet{Chachan2023}, the abundance ratios for Si/H, C/H, and O/H are given relative to their stellar values.}
    \label{fig:chachan_scenarios}
\end{figure*}

\subsection{Linking the refractory abundance to planet formation}

To asses the link between the present-day atmospheric abundance and the formation history of TWA\,5\,B, we adopt the abundance ratio tracks from \citet{Chachan2023}, which model the volatile-to-refractory ratios produced by core-accretion in different regions of a protoplanetary disc. The observed volatile-to-refractory ratios of TWA\,5\,B are consistent, at 2$\sigma$ confidence, with two possible formation pathways: i) formation just within the CO snowline and ii) beyond the CO snowline. These core accretion scenarios are broadly consistent with formation in the outer region of the disc, in line with the present-day location of TWA\,5\,B. In the outer disc the majority of volatiles would have been accreted as solids, matching the observed super-solar Si/H, which indicates substantial accretion of silicate-rich solids. This solid-rich accretion is additionally reflected in the approximately solar C/Si and O/Si ratios, as beyond the CO snowline carbon and oxygen bearing species are accreted as ices and will therefore have C/Si and O/Si ratios that match the composition of the host star \citep{Pelletier2025}.

Previous inferences using the models of \citet{Chachan2023} and volatile-to-refractory ratios for the ultra-hot Jupiters WASP-121 b and WASP-189 b have implied a range of possible formation locations. For WASP-121 b \citet{Smith2024} measure atmospheric abundances that favour formation between the soot (refractory carbon) line and H$_2$O snowline, but cannot exclude far out formation between the H$_2$O and CO snowlines or beyond the CO snowline, while \citep{Pelletier2025} favour formation near the CO snowline. Similarly, volatile-to-refractory measurements for WASP-189 b favour close in formation, but cannot rule out formation beyond the CO snowline. Significant planetary migration is required to reconcile the inferred formation locations of these short period planets with their present-day orbits. In contrast, for the directly imaged companion TWA\,5\,B, the favoured scenario for core accretion formation is localised to the outer disc beyond the CO snowline, consistent with the present-day location of the companion. 

However, forming the high mass of TWA\,5\,B (25~$\pm$~5~M$_J$) through core accretion is challenging to reconcile with the predicted disc mass of the TWA\,5 system. The disc mass of a single or close binary system is observed to be directly proportional to the host star mass, with a typical disc to star mass ratio M$_{\text{disc}}$/M$_{\text{star}}$ $\approx$ 0.4 $\pm$ 0.2$\%$ \citep{Andrews2013}. For the total 0.9 $\pm$ 0.1 M$_{\odot}$ mass of the TWA\,5\,Aab binary stars \citep{Kohler2013}, we would therefore predict a total protoplanetary disc mass of $\sim$ 4 $\pm$ 2 M$_{\text{J}}$, far below the requirement to produce TWA\,5\,B through core accretion.

Given the wide orbit (a $\approx$ 80 AU) and high mass (25~$\pm$~5~M$_J$) of TWA\,5\,B, we might expect that star-like formation processes such as gravitational instability to be the favoured formation scenarios \citep{Boss1997, Durisen2007}. Gravitational instability, which proceeds via the rapid collapse of a local region of the protostellar disc, is predicted to form a gaseous envelope that initially inherits the abundance ratios of the host star. However, the engulfment of solids during the collapse of the disc fragment, post-formation accretion of planetesimals from the disc, or preferential accretion of gas, ices, or silicates, can significantly alter the present-day metal enrichment and elemental abundance ratios of the body \citep{Helled2010,Helled2011, Boss2026}. This post-collapse enrichment is dependent on the presence of sufficient material in the remaining protoplanetary disc. Adopting a generous upper limit of 50 M$_{\text{J}}$ for a very massive disc, assuming a 1$\%$ dust-to-gas ratio, and using a power-law for the surface density ($\sigma$ = $\sigma_0$ (a/5 AU)$^{\alpha}$, with $\alpha$ = 1.00; following \citealt{Helled2009, GRAVITY2020}), the solid surface density at the present-day location of TWA\,5\,B ($\sim$ 80 AU) is $\sigma(80 \text{AU}) \approx 0.413$ g/cm$^2$. The orbital period of TWA\,5\,B is $\approx$ 800 yrs, resulting in an orbital frequency $\Omega \approx$ 4 $\times$ 10$^{-11}$. Adopting an effective capture radius of 5 $\times$ 10$^{12}$ cm, following \citep{Helled2006, GRAVITY2020}, provides 

\begin{equation}
M_{\text{accreted}} \approx 7.0 \times 10^{-6} M_{\oplus} \times t_{\text{collapse}} [\text{years}].
\end{equation}

Therefore in a 10$^3$ year collapse timespan \citep{Helled2006}, only a very minor quantity of solids can be accreted during a gravitational instability formation ($\sim$ 0.005 $M_{\oplus}$). Continued planetesimal accretion across the 10 $\pm$ 3 Myr lifespan of TWA\,5\,B, assuming a constant accretion rate and an undepletable reservoir of planetesimals at the companion location, would yield a maximum accretable mass of M$_{\text{accreted}} \approx$ 65 $M_{\oplus}$. For TWA\,5\,B, with a mass of 25~$\pm$~5~M$_J$, a significant deviation of the global atmospheric abundance ratios from their stellar values would require substantial accretion of material post-collapse. We can estimate the mass of silicates that must be accreted to produce the observed enrichment in silicon of $\sim$ 25 times the stellar value, through the following relation. The additional silicon required to be accreted (M$_{\text{Si, acc}}$) is given by: 

\begin{equation}
M_{\mathrm{Si, acc}} = 25~X_{\mathrm{Si,\star}}~M_{\mathrm{envelope}}
\end{equation}

where the stellar mass fraction of silicon for TWA\,5 A is X$_{\mathrm{Si, \star}}$ = 9.1 $\times$ 10$^{-5}$. If we assume that the entire envelope is convective and well mixed, the mass of the envelope for enrichment ($M_{\text{envelope}}$) is the total mass of TWA\,5\,B (25~$\pm$~5~M$_J$), leading to an accreted mass of silicon $M_{\mathrm{Si, acc}}$ = 20 M$_{\oplus}$. If considering the star to have solar abundances, this leads to a requirement of $M_{\mathrm{Si, acc}}$ = 4 M$_{\oplus}$. These masses of accreted silicon are in excess of what can be accreted during the pre-collapse phase of gravitational instability formation, determined above to be $\sim$ 0.005 $M_{\oplus}$, but appear consistent with the maximum possible refractory material accreted post-formation, estimated to be $\approx$ 65 $M_{\oplus}$. Therefore a gravitational instability formation pathway for TWA\,5\,B, in which TWA\,5\,B has inherited the primordial competition of the protostellar disc, followed by the enrichment of silicates appears, to first order, to be able to match the observed enrichment of silicon. This does require extensive late stage disc accretion of silicate material, and TWA\,5\,B does indeed show signatures of H$\alpha$ emission \citep{Neuhauser2000}, and X-ray emission \citep{Tsuboi2003}, potentially indicative of ongoing accretion.  

Collectively, the measured atmospheric abundances support a formation in the outer regions of the disc, consistent with the present-day location and mass of TWA\,5\,B. The inclusion of refractory elements in this inference provides additional confidence in this conclusion. When comparing to the core-accretion models, the use of refractory elements to supplement the C/O ratio more precisely constrains the permitted formation locations in the disc. When considering gravitational instability formation for TWA\,5\,B, the measurement of silicon directly probes the extent of silicate accretion. For further inference on planet formation, however,  population studies are required, using many complementary tracers, both atmospheric tracers (C/O, M/H, O/Si, C/Si, Si/H, $^{12}$C/$^{13}$C) and fundamental planetary parameters (e.g. T$_{\text{eff}}$, M$_{\text{p}}$), interpreted within the planetary system context.

\subsection{SiO as a tracer of the onset of cloud condensation across the hot giant planet population}

The condensation of magnesium silicate cloud species is a dominant process that determines the properties of the population of hot directly imaged planets and free-floating brown dwarfs. Identifying the cloud condensation temperatures, and consequently the prevalent cloud species across the directly imaged planet population, is vital to understanding the chemical and thermal evolution of these objects. Furthermore, uncertainties surrounding the exact cloud species forming on hot giant planets currently limit our interpretation of observed atmospheric spectra. The condensation of different silicate condensates e.g. MgSiO$_3$ (enstatite), Mg$_2$SiO$_4$ (forsterite), or SiO$_2$ (quartz) sequester different amounts of oxygen and metals, biasing inferred C/O and metallicity measurements from all observed spectra \citep{Lacy2020}.

The inference of specific condensates in young internally heated planetary atmospheres has previously been carried out through observations of the 8--11 $\mu$m silicate absorption feature in the isolated brown dwarf population with Spitzer and JWST, both at a population level \citep{Suarez2022, Lam2026}, and through detailed studies of the cloud species and properties of individual objects \citep[e.g.][]{Burningham2021, Vos2023}. Recently, the use of JWST/MIRI to target the 8--11 $\mu$m silicate feature in planetary mass objects has presented significant diversity in the detected cloud species. \citet{Molliere2025} detect cloud features in the isolated planetary-mass brown dwarf PSO J318.5338-22.8603, which are best fit with < 0.1 $\mu$m amorphous SiO grains. For the directly imaged planet YSES 1 c, \citet{Hoch2025} similarly observe < 0.1 $\mu$m sized cloud particles, but the 8--11 $\mu$m silicate feature is best fit by either iron-enriched pyroxene (Mg$_{0.8}$Fe$_{0.2}$SiO$_3$) or a combination of amorphous MgSiO$_3$ and Mg$_2$SiO$_4$. These observations are in contrast to JWST/MIRI mid-infrared transmission observations of short period giant planets, which suggest SiO$_2$ clouds are dominant on the highly irradiated hot Jupiters WASP-17 b and HD 189733 b \citep{Grant2023, Inglis2024}.

\begin{figure}
    \includegraphics[width=\columnwidth]{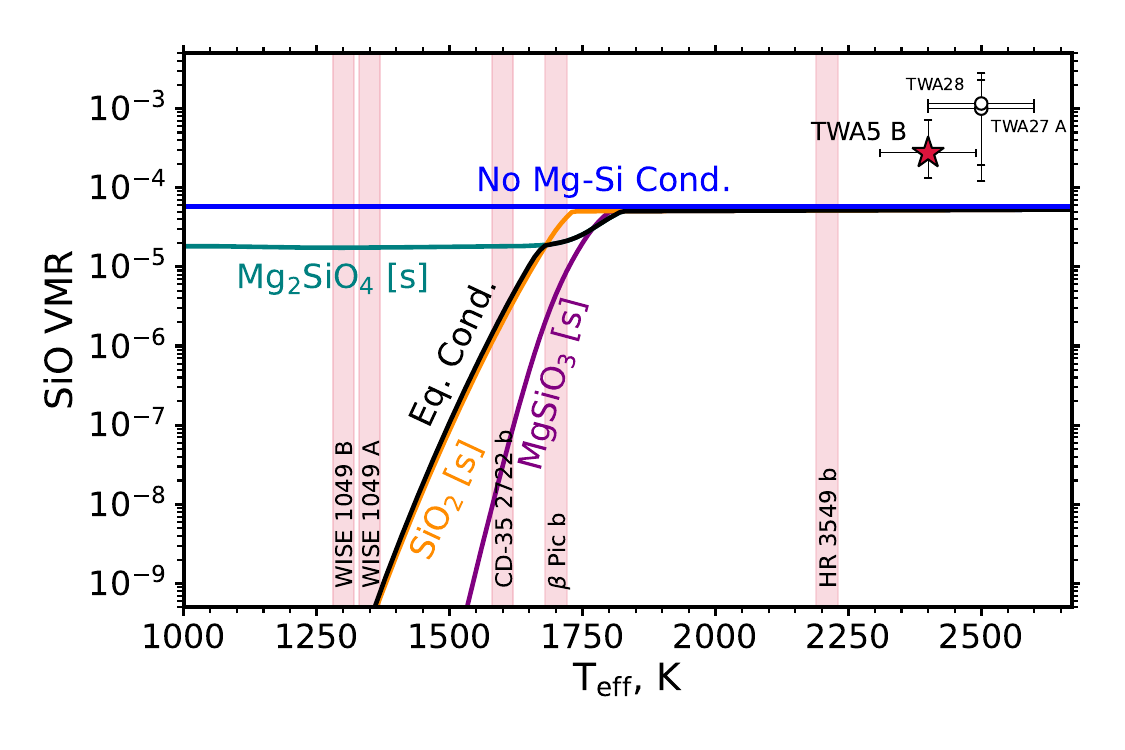}
    \caption{Predicted gas phase SiO abundance in the atmospheres of directly imaged planets, under the influence of different magnesium-silicate condensates, generated with \textsc{FastChemCond} rain out condensation \citep{Kitzmann2024}. The model tracks present the expected gas phase abundance of SiO for sub-stellar objects of different temperatures, with each track representing an atmosphere in which only a single cloud species can condense. For example, if MgSiO$_3$ clouds preferentially form in sub-stellar atmospheres we expect the SiO abundance to be rapidly depleted at $\sim$1800 K following the MgSiO$_3$ track (purple), in comparison to the case in which Mg$_2$SiO$_4$ are the dominant condensate, in which the observed SiO abundance will not drop below 10$^{-5}$ VMR (teal). The `Eq. Cond.' track (black) denotes the equilibrium chemistry condensation prediction. The existing SiO measurements from isolated brown dwarfs are plotted \citep{GonzalezPicos2025b}, alongside the abundance measurement of SiO in TWA\,5\,B presented in this work. The remaining targets in the CRIMSON survey (highlighted in red) will probe cooler temperatures, providing an opportunity to test the cloud species forming in this population.}
    \label{fig:cond}
\end{figure}

To move beyond single target studies and perform an analysis of cloud condensation on a population level, we require observables which can be readily detected and reliably linked to cloud formation. The abundance of gaseous SiO in giant planets provides a unique diagnostic of cloud condensation, and the retrieval of SiO abundances using CRIRES+ \textit{M}-band offers a mechanism to reliably measure this abundance across the hot-giant planet population. SiO is abundant in giant planets in chemical equilibrium with a VMR $\approx$ 10$^{-4}$, as detected for TWA\,5\,B, but in the presence of magnesium-silicate cloud decks the SiO abundance is rapidly depleted in proportion to the level of condensation as the gaseous Si is sequestered into clouds \citep{Visscher2010}.

As gaseous SiO contains almost the entire atmospheric silicon abundance, and is depleted only through condensation in hot giant planets, it is an ideal tracer for determining the onset of different cloud species. In Figure \ref{fig:cond} we demonstrate simulations using \textsc{FastChemCond} \citep{Kitzmann2024}, of the predicted gas phase SiO abundance across a range of planetary temperatures, considering the formation of a range of different condensates. These simulations adopt internally heated TP profiles \citep{Guillot2010} and rainout condensation, with the predicted observable SiO gas phase abundance calculated by integrating over the pressure levels probed with CRIRES+ \textit{M}-band spectroscopy at 4 $\mu$m (-2 $\lesssim$ log(P [bar]) $\lesssim$ 0). A fixed log(g) = 4 is assumed, and equilibrium chemistry applied with a solar metallicity. To produce the SiO gas phase abundance predictions for the individual molecules, \textsc{FastChemCond} is fixed to consider only the single condensate in question (MgSiO$_3$, Mg$_2$SiO$_4$, and SiO$_2$). The model labelled `Eq. Cond.' contains all magnesium-silicate cloud species, and denotes a prediction that is thermodynamically favoured for substellar objects with solar Mg/Si ratios; when moving from higher to lower temperatures we see the onset of Mg$_2$SiO$_4$ condensation at T $\approx$ 1800 K, followed by the rapid depletion of gas phase SiO from the condensation of SiO$_2$ at T $\approx$ 1650 K. For the directly imaged planets, the abundance of gas phase SiO has a dependence on metallicity, and shows variations as a function of surface gravity due to the impact of the surface gravity on the TP profile, requiring accurate measurements of M/H and the TP profile. 

We propose that by surveying the SiO abundance across the population of directly imaged planets, the condensation temperature of cloud species can be determined, thus determining the dominant condensate. Of particular interest is the discrepancy in the SiO gas phase abundance between Mg$_2$SiO$_4$ and MgSiO$_3$ condensation, which is readily discernible across the population. Mg$_2$SiO$_4$ condensation is limited by the availability of magnesium rather than silicon \citep{Calamari2024}, and therefore for atmospheres with solar Mg/Si compositions it will not deplete the observed SiO abundance below 10$^{-5}$ VMR (Figure \ref{fig:cond}). The existing abundance measurements of gas phase SiO in imaged objects are plotted, for the isolated brown dwarfs TWA\,28 and TWA\,27 A \citep{GonzalezPicos2025b}, and the abundance measurement of SiO in the companion TWA\,5\,B presented in this work. These targets fall in the regime in which SiO is not expected to be depleted by clouds, and pushing to cooler temperature objects with the remaining CRIMSON survey targets will be essential to probe this trend.

The level of gas phase SiO depletion measured across a population has implications for theoretical models of cloud formation, and can help determine whether sedimentation and rainout drive cloud formation (e.g. \citealt{Ackerman2001}), or whether cloud seed particles form deep in the atmosphere and are transported upward (e.g. \citealt{Helling2006}). The present discussion has been limited to rainout condensation \citep{Ackerman2001}, in which particles instantaneously condense from the gas phase based on the local thermodynamic equilibrium, uniformly depleting the atmosphere above a cloud deck of specific gas phase species, in this case SiO. However, in the \citet{Helling2006} model of kinetic cloud formation, the depletion of cloud precursor species above a cloud deck arises from kinetically controlled surface reactions on the lofted cloud seed nuclei. As a consequence, the level of gas phase SiO depletion is dependent on the relative efficiencies of atmospheric mixing compared with the efficiency of condensate grain growth and nucleation. The depletion is also predicted to show a less abrupt profile of depletion with altitude, with the atmosphere immediately above the cloud deck showing the largest depletion of gas phase SiO. These contrasting predictions of SiO abundance from cloud formation models, either the total depletion of SiO above the cloud deck from rainout condensation \citep{Ackerman2001}, or the more partial depletion predicted by kinetic cloud formation \citep{Helling2006, Helling2008}, could be directly tested by tracing the depletion of cloud precursor species such as SiO across the giant planet population.

The inference of magnesium-silicate cloud properties through the depletion of silicon-bearing gases has a historical Solar System precedent. The absence of observable silane (SiH$_4$), the dominant carrier of silicon in cooler atmospheres, in the upper atmospheres of Jupiter and Saturn is due to the formation of deep magnesium silicate cloud decks, depleting the upper atmosphere of refractory Si \citep{Fegley1994}. An equivalent test for the presence of clouds in M/L dwarfs using gas phase silicate species has previously been proposed (e.g. \citealt{Lodders2006, Visscher2010}), and here we identify and demonstrate gaseous SiO to be a readily observable and direct tracer of cloud formation.

\section{Conclusions}
\label{sec:Conclusions}

We have characterised the atmosphere of the directly imaged companion TWA\,5\,B using high-resolution spectroscopy in the CRIRES+ \textit{M}-band, with 2.2 hours of exposure time on source. We conclude the following:

\begin{enumerate}[label=\roman*., leftmargin=*, align=left, itemsep=0.8em]
    \item We implement a customised extraction routine to mitigate the sampling systematic seen when extracting spectra at the spatial pixel scale of CRIRES+. We identify that this systematic may impact all echelle slit spectrographs when spectra are extracted at the spatial pixel scale, including next-generation instruments such as METIS/ELT.
    
    \item Cross-correlating with model templates we detect gaseous CO (S/N = 9.1), H$_2$O (S/N = 18.8), and SiO (S/N = 7.5) in TWA\,5\,B. The detection of SiO marks the first direct detection of gas phase silicon chemistry in a directly imaged companion. Building on the first use of \textit{M}-band HRS in \citet{Parker2024}, this work represents a high S/N demonstration of the capabilities of CRIRES+ \textit{M}-band, and a preview of the scientific potential of METIS/ELT.
    
    \item With a free-chemistry atmospheric retrieval using the HyDRA framework to infer the directly imaged companion properties (chemical abundances, atmospheric dynamics, and temperature-pressure profile) we measure a rotational broadening of the TWA\,5\,B spectrum, retrieving a projected rotational velocity of vsin(i) = 14.19$^{+0.88}_{-0.79}$ km/s, consistent with literature measurements.
    
    \item We obtain a well constrained and high abundance of gaseous SiO, log(SiO) = $-3.56^{+0.42}_{-0.32}$ VMR, confirming theoretical predictions of SiO as the dominant gas phase carrier of silicon in hot giant exoplanets and sub-stellar companions. This high SiO abundance indicates the lack of significant sequestration of gas phase silicon into condensates, and we thus determine that the atmosphere is free from silicate clouds at or deeper than the -2 $\lesssim$ log(P [bar]) $\lesssim$ 0 pressures probed in these observations. The lack of evidence for cloud opacity at lower pressures supports the absence of silicate clouds across the entire atmosphere. 
    
    \item The direct sensitivity of the gas phase SiO abundance to the condensation of silicate clouds provides a compelling litmus test for inferring the condensation properties of cloud species in hot giant planet atmospheres. We model the predicted SiO abundance with temperature for a range of prominent condensates, and demonstrate that the gas phase SiO VMR is sensitive to the exact composition of the condensate forming (e.g. Mg$_2$SiO$_4$ or MgSiO$_3$), alongside the temperature of the planetary atmosphere. Relatively inexpensive observations of SiO with high-resolution \textit{M}-band spectroscopy therefore provide a pathway to determine cloud compositions across the giant planet and isolated brown dwarf populations, without the requirement of space-based JWST spectra.
    
    \item For TWA\,5\,B, the lack of strong evidence for sequestration of silicon into clouds implies that the near entirety of the refractory silicon above the photosphere is contained within gaseous SiO. Therefore, we are directly probing the refractory enrichment of the atmosphere of TWA\,5\,B. We calculate marginally sub-stellar volatile-to-refractory ratios of O/Si and C/Si, but a super-stellar Si/H ratio, [Si/H]$_{\star}$ = $1.41^{+0.42}_{-0.32}$, suggesting significant enrichment of silicon in the atmosphere of TWA\,5\,B relative to the host star.
    
    \item When considering the high mass (25~$\pm$~5~M$_J$) and wide orbit (a $\approx$ 80 AU) of TWA\,5\,B, gravitational instability is the favoured formation pathway but would require substantial accretion of silicate-rich solids post-collapse to match the observed silicon enrichment. Conversely, atmospheric volatile-to-refractory ratios, when evaluated in conjunction with the atmospheric metallicity and C/O ratio, are also consistent with formation through core-accretion in the outer regions of the disc near or beyond the CO snowline, but a core-accretion formation for TWA\,5\,B is in tension with the predicted disc mass required to form a 25~$\pm$~5~M$_J$ object at the observed separation.

\end{enumerate}

\section*{Acknowledgements}

We thank the referee for their helpful comments that improved the quality of the manuscript. We thank Jonathan Fortney, Ravit Helled, and Paul Mollière for helpful discussions on the role of silicon in giant planet atmospheres. We additionally thank Mario van den Ancker and Sam de Regt for discussion of observation planning for the CRIMSON survey targets. LTP and JLB acknowledge funding from the European Research Council (ERC) under the European Union’s Horizon 2020 research and innovation program under grant agreement No 805445. JLB further acknowledges the support of the Leverhulme Trust via the Philip Leverhulme Physics Prize. This study is based on observations collected at the European Organisation for Astronomical Research in the Southern Hemisphere under ESO programme 114.27LL. This research has made use of the NASA Exoplanet Archive, which is operated by the California Institute of Technology, under contract with the National Aeronautics and Space Administration under the Exoplanet Exploration Program. This research has made use of NASA’s Astrophysics Data System Bibliographic Services and the SIMBAD database, operated at CDS, Strasbourg, France. This research made use of SAOImageDS9, a tool for data visualization supported by the Chandra X-ray Science Center (CXC) and the High Energy Astrophysics Science Archive Center (HEASARC) with support from the JWST Mission office at the Space Telescope Science Institute for 3D visualization \citep{Joye2003}. This work made use of the whereistheplanet prediction tool \citep{Wang2021b}. This work has made use of the Python programming language\footnote{\url{https://www.python.org/}}, in particular packages including NumPy \citep{Harris2020}, SciPy \citep{Virtanen2020}, Matplotlib \citep{Hunter2007}, and Astropy \citep{Astropy2013,Astropy2018,Astropy2022}.

\section*{Data Availability}

The raw data used in this study is available for download from the ESO Data Archive under Programme ID 114.27LL. Processed data products and models are available on reasonable request to the corresponding author.


\bibliographystyle{mnras}
\bibliography{TWA5}


\appendix
\label{sec:Appendix}

\section{Full atmospheric retrieval corner plot}
\label{sec:App_cornor_plot}

\begin{figure*}
    \begin{center}
    \includegraphics[width=1.98\columnwidth]{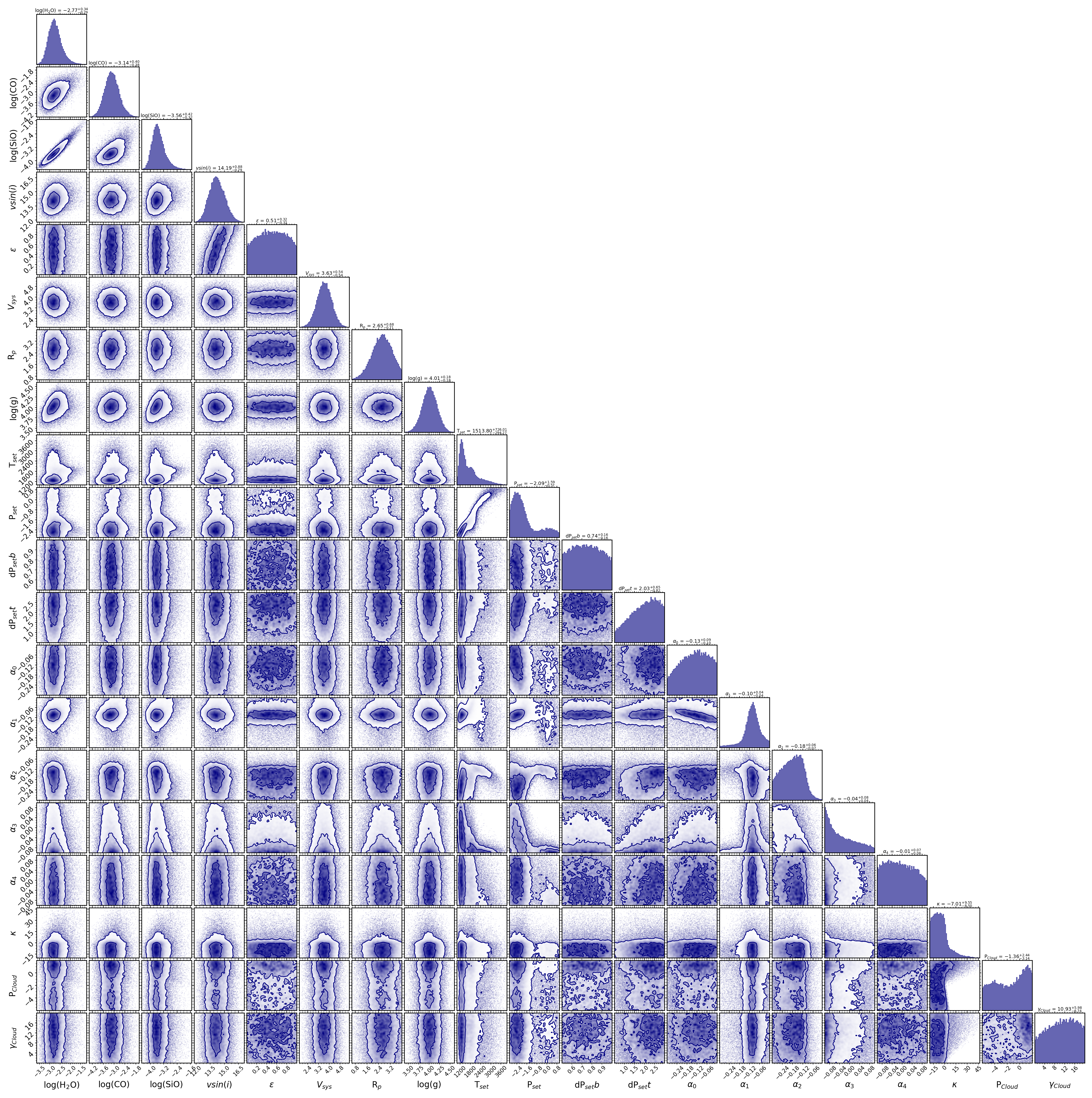}
    \caption{Full corner plot from the free chemistry retrieval.}
    \label{fig:Big_corner_plot}
    \end{center}
\end{figure*}


\bsp	
\label{lastpage}
\end{document}